\documentclass[lettersize,journal]{IEEEtran}
\usepackage{amsmath,amsfonts}
\usepackage{algorithmic}
\usepackage{algorithm}
\usepackage{array}
\usepackage[caption=false,font=normalsize,labelfont=sf,textfont=sf]{subfig}
\usepackage{textcomp}
\usepackage{stfloats}
\usepackage{url}
\usepackage{verbatim}
\usepackage{graphicx}
\usepackage{cite}
\usepackage{amssymb}
\usepackage{booktabs}
\usepackage{multirow}
\usepackage{xcolor}
\begin{document}

% ===== TITLE & AUTHORS =====
\title{Neighbor-Consistent Neural Filters for Robust Personal Sound Zones Under Localization Uncertainty}

\author{
Hao~Jiang, Edgar~Choueiri
\thanks{
The authors are with the 3D Audio and Applied Acoustics (3D3A) Laboratory,
Princeton University, Princeton, NJ 08544 USA 
(Corresponding author: Hao Jiang; e-mail: hj3737@princeton.edu; choueiri@princeton.edu).
}%
}

% The paper headers
\markboth{IEEE Transactions on Audio, Speech, and Language Processing,~Vol.~XX,~XXXX}%
{Jiang and Choueiri: Neighbor-Consistent Neural Filters for Robust Personal Sound Zones Under Localization Uncertainty}

\maketitle

% ===== ABSTRACT =====
\begin{abstract}
Coordinate-conditioned neural networks can generate head-tracked personal sound zone (PSZ) loudspeaker filters in real time, but they are sensitive to localization uncertainty. Small fluctuations in the estimated listener coordinates (e.g., coordinate estimation variance stemming from optical distortion, momentary occlusions, or algorithmic tracking jitter) may produce large changes in the generated filters even when listeners are physically stationary. This paper proposes neighbor-consistent neural filters that regularize the coordinate-to-filter mapping by penalizing filter differences at randomly perturbed neighboring coordinates during training. To objectively evaluate robustness against tracking noise, we introduce a decoupled protocol that fixes the acoustic transfer functions at a physical anchor while perturbing only the coordinate inputs used for filter generation to emulate head-tracking noise. Isolation quality and local stability are assessed using objective metrics, including neighborhood median and lower-tail statistics of inter-zone and inter-program isolation, as well as spatial variation rates that quantify metric sensitivity within a coordinate neighborhood. In simulation with a split-band woofer--tweeter system and 25 randomly sampled anchor positions, neighbor consistency reduces the root-mean-square (RMS) variation rate by up to 55.9\% in the woofer band and 30.3\% in the tweeter band, while largely preserving neighborhood isolation quality and improving lower-tail robustness. In in-situ measurements using a 24-driver array and two stationary head-and-torso simulators, the proposed regularization improves worst-case neighborhood isolation by up to 16.9\% and reduces spatial variation rates by up to 61.8\%, with improvements observed for both listeners under coordinate-input perturbations. These objective results demonstrate that neighbor-consistency regularization is effective at stabilizing PSZ rendering under localization uncertainty.
\end{abstract}

% ===== INDEX TERMS =====

%%%%%%%%%%%%%%%%%%%%%%%%%%%%%%%%%%%%%%%%%%%%%%%%%%%%%%%%%%%%%%
%%%%%%%%%%%%%%%%%%%%      MAIN TEXT      %%%%%%%%%%%%%%%%%%%%%%
%%%%%%%%%%%%%%%%%%%%%%%%%%%%%%%%%%%%%%%%%%%%%%%%%%%%%%%%%%%%%%
\section{Introduction}
\label{sec:intro}

\IEEEPARstart{P}{ersonal} Sound Zones (PSZs) aim to deliver multiple, independent audio programs to different listeners who share the same acoustic space without headphones. A typical PSZ system creates one or more bright zones where a target program is reproduced, while maintaining dark zones where that program is suppressed, using a loudspeaker array and pre-designed or adaptive filters. The concept dates back to early demonstrations \cite{druyvesteyn1997personalsound} of ``personal sound'' and has since developed into a broad research area with applications in vehicles, homes, and shared spaces \cite{vindrola2021car,betlehem2015psz,wallace2020speechprivacy,jacobsen2023living}. PSZs can be viewed as a specialized instance of multi-point sound field control, building on foundational work in sound field synthesis and reproduction with loudspeaker arrays \cite{ahrens2010sfr,gerzon1992metatheory,berkhout1993wfs,ward2001planewave,poletti2005spherical}.

Most PSZ design methods optimize a trade-off between reproducing the target program accurately in the bright zone and suppressing it in the dark zone. Acoustic Contrast Control (ACC) maximizes the energy ratio between zones and is widely used due to its simplicity and strong isolation in many settings \cite{choi2002acc,chang2009realization}. Pressure Matching (PM) minimizes a reproduction error, typically improving fidelity at the cost of reduced contrast, and many variants have been proposed to better balance distortion and isolation \cite{wu2011multizone,moles2022wpmwindow}. Practical systems further face bandwidth-dependent actuator limits and placement constraints, which motivates split-band or multi-layer architectures \cite{cheer2013carcabin,shin2014duallayer}. For example, separating low- and high-frequency control (e.g., woofer--tweeter or dual-array setups) can reduce hardware demands while leveraging frequency-dependent controllability.

Robustness is another long-standing concern in sound zone control. Performance can degrade under reverberation, acoustic transfer function (ATF) mismatch, or other modeling errors, motivating diagonal loading, regularization, and robust optimization \cite{elliott2012robust,doclo2003robustbeamformers,bai2014montecarlo}. Robust formulations have also incorporated acoustic modeling, reduced in-situ measurements, and explicit uncertainty constraints on ATFs \cite{zhu2017robustaccmodel,zhang2023cgmmrpm}. Related work has studied robustness for directional or locally oriented reproduction objectives, again showing that good nominal performance does not necessarily imply stable performance under uncertainty \cite{zhu2017robustrepro}. In parallel, efficient implementations continue to broaden the design space, including subband and time--frequency approaches that trade off controllability, latency, and computation \cite{moles2020subband,tang2025stft}.

Recently, learning-based approaches have been explored to replace or augment optimization-based filter design. Early work investigated neural networks for PSZ filter design \cite{pepe2022neural}. Subsequent efforts developed coordinate-conditioned models that map listener positions directly to filter coefficients, enabling fast adaptation for dynamic or head-tracked rendering \cite{qiao2025sannpsz}. Building on this adaptive framework, binaural-focused neural models have also been introduced to specifically optimize head-tracked binaural rendering \cite{jiang2026bsann}. Once trained, these models provide rapid filter inference, avoiding the repeated optimization otherwise required at run time.

This paper focuses on a robustness issue that is especially important for coordinate-conditioned filter generation: localization uncertainty. In practice, a listener may remain physically stationary, but the estimated position used by the system (from head tracking or localization sensors) can fluctuate due to finite sensor resolution, optical distortion, noise in the sensor data, or algorithmic jitter in the pose estimation. We collectively refer to such estimation-induced coordinate perturbations as \emph{tracking noise} throughout this paper. If the coordinate-to-filter mapping is sensitive, small coordinate errors can cause noticeable changes in the reproduced field even without any physical motion. This uncertainty is different from ATF mismatch: ATF mismatch changes the underlying acoustic model, whereas localization uncertainty perturbs only the coordinate conditioning signal while the physical acoustics at the listener position remain fixed. Standard PSZ evaluations typically report absolute isolation and fidelity at a limited number of anchor points; however, such metrics do not directly characterize stability under coordinate perturbations. While recent work has clarified principled isolation and interference metrics for PSZs \cite{qiao2022metrics}, explicit stability metrics are still needed when the system depends on noisy coordinate data.

We propose neighbor-consistent neural filters for PSZ rendering under localization uncertainty. The core idea is to regularize the coordinate-conditioned filter generator so that spatially neighboring input coordinates produce similar filters, thereby yielding more stable sound fields at a fixed physical listener position. This approach can be viewed as a form of consistency regularization, a common strategy in machine learning to reduce sensitivity to input perturbations \cite{laine2017temporal,tarvainen2017mean,miyato2018vat,sohn2020fixmatch}. Similar neighborhood-based constraints have also shown success in enforcing spatial continuity for learning-based HRTF upsampling~\cite{hu2025hrtfformer}. In our context, the perturbation has a direct physical meaning: it represents uncertainty in the coordinate input to the system.

We also introduce an evaluation protocol that separates physical acoustics from noisy coordinate conditioning. Specifically, we evaluate using acoustic transfer functions (ATFs) for a fixed physical listener position, and perturb only the input coordinates provided to the filter generator. This isolates the effect of localization uncertainty from true listener motion. We report both (i) quality-oriented metrics that characterize isolation performance and (ii) stability-oriented metrics that quantify how rapidly performance changes within a small coordinate neighborhood, thereby making the robustness--quality trade-off explicit.

The main contributions of this work are:
\begin{itemize}
    \item We propose a neighbor-consistency regularization for coordinate-conditioned neural PSZ filter generation, improving spatial stability under localization uncertainty.
    \item We introduce a decoupled evaluation protocol that isolates tracking noise from physical-displacement-induced acoustic transfer variations by keeping the acoustic transfer functions fixed at the true physical configuration while perturbing the coordinate input used for filter generation.
    \item We report stability-oriented metrics based on neighborhood step statistics normalized by spatial displacement, alongside standard PSZ isolation metrics defined in~\cite{qiao2022metrics}, and quantify the associated performance trade-offs.
    \item We validate the proposed approach in both simulation and in-situ measurements using a 24-driver split-band woofer--tweeter loudspeaker system and two stationary head-and-torso simulators (HATS).
\end{itemize}

The remainder of this paper is organized as follows. Section~\ref{sec:method} presents the proposed neighbor-consistent training objective and filter-generation framework. Section~\ref{sec:eval} describes the decoupled evaluation protocol, alongside the isolation quality and spatial stability metrics. Section~\ref{sec:sim} reports simulation results, Section~\ref{sec:meas} reports measurement validation, and Section~\ref{sec:concl} concludes. A hyperparameter sensitivity study is provided in Appendix~\ref{app:hparam}.

%%%%%%%%%%%%%%%%%%%%%%%%%%%%%%%%%%%%%%%%%%%%%%%%%%%%%%%%%%%%%%
%%%%%%%%%%%%%%%%%%%%     Method     %%%%%%%%%%%%%%%%%%%%%%%
%%%%%%%%%%%%%%%%%%%%%%%%%%%%%%%%%%%%%%%%%%%%%%%%%%%%%%%%%%%%%%

\section{Neighbor-Consistent Neural Filter Generation}
\label{sec:method}

\begin{figure*}[t]
  \centering
  \includegraphics[width=\textwidth]{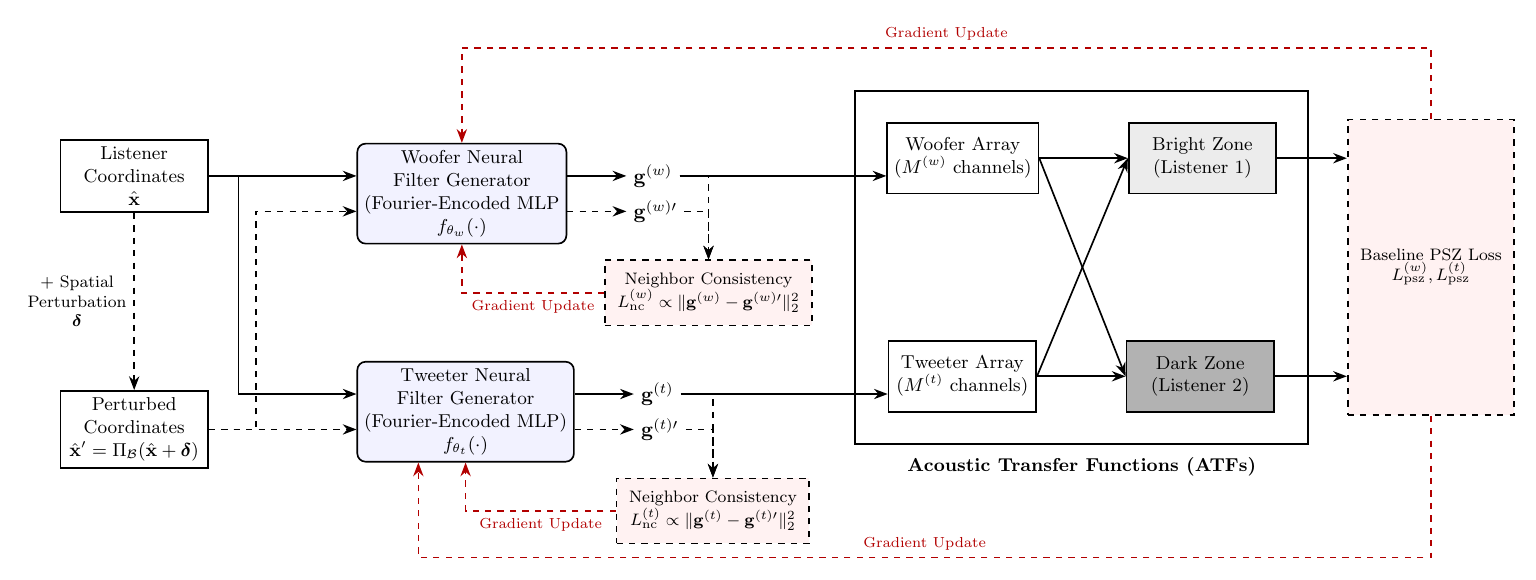}
  \caption{Coordinate-conditioned neural PSZ filter generation for a split-band (woofer--tweeter) system using two independently trained models. The woofer model $f_{\theta_w}$ and tweeter model $f_{\theta_t}$ map estimated listener coordinates to band-specific FIR filters. Neighbor-consistency regularization is applied during training by penalizing filter differences for perturbed input coordinates.}
  \label{fig:system_overview}
\end{figure*}

This section presents the coordinate-conditioned neural filter generators and the proposed neighbor-consistency regularization. The objective is to reduce filter sensitivity to coordinate-input perturbations arising from head-tracking or localization uncertainty. An overview of the split-band pipeline and the two independently trained generators is shown in Fig.~\ref{fig:system_overview}.

% -------------------------------------------------------------------------

\subsection{Split-band system model}
\label{subsec:signal_model}

We consider a loudspeaker system composed of two driver groups: woofers for low-frequency reproduction and tweeters for high-frequency reproduction. In this work, the woofer model controls the 100--2000~Hz band, while the tweeter model controls the 2--20~kHz band. This split-band structure is common in practical personal-audio systems and is closely related to multi-layer sound zone implementations \cite{cheer2013carcabin,shin2014duallayer}. Let $M^{(w)}$ and $M^{(t)}$ denote the number of woofer and tweeter channels, respectively. We assume $K$ audio programs are delivered to $K$ listeners/zones; throughout this work we focus on $K=2$.

Let $s_{k,c}[n]$ denote the discrete-time target audio program for zone $k$, where $c \in \{L, R\}$ represents the independent left and right audio channels intended for the corresponding ears of the listener. For each band $b\in\{w,t\}$, loudspeaker channel $m\in\{1,\dots,M^{(b)}\}$, program $k\in\{1,\dots,K\}$, and audio channel $c$, we design an FIR filter $g^{(b)}_{m,k,c}[n]$ of length $L^{(b)}$. The drive signal for loudspeaker $m$ in band $b$ is
\begin{equation}
v^{(b)}_{m}[n] = \sum_{k=1}^{K} \sum_{c \in \{L,R\}} (g^{(b)}_{m,k,c} * s_{k,c})[n],
\label{eq:drive_signal}
\end{equation}
where $*$ denotes convolution. Let $N_e$ denote the number of spatial control points per ear. We define $\mathbf{h}^{(b)}_{k,c,m}[n] \in \mathbb{R}^{N_e}$ as the vector of discrete-time room impulse responses (RIRs) from channel $(m,b)$ to the $N_e$ control points at ear $c$ of listener $k$. The reproduced pressure vector at ear $c$ of listener $k$ is given by
\begin{equation}
\mathbf{p}_{k,c}[n] = \sum_{b\in\{w,t\}} \sum_{m=1}^{M^{(b)}} \big(\mathbf{h}^{(b)}_{k,c,m} * v^{(b)}_{m}\big)[n],
\label{eq:pressure_sum}
\end{equation}
where the convolution is applied element-wise. 

For compact notation, we concatenate all FIR filter coefficients for band $b$ into a single column vector
\begin{equation}
\mathbf{g}^{(b)} \in \mathbb{R}^{D^{(b)}}, \qquad
D^{(b)} = 2\, K\, M^{(b)} L^{(b)},
\label{eq:filter_vector_band}
\end{equation}
which serves as the target output space for the neural filter generators.

\subsection{Coordinate-conditioned neural filter generators}
\label{subsec:two_generators}

Let $\mathbf{x}_k \in \mathbb{R}^{d}$ denote the center-of-head position vector of listener $k$ (typically $d=2$ on a horizontal plane). The corresponding spatial control points for the left and right ears, as required by the physical model in Section~\ref{subsec:signal_model}, are geometrically derived from $\mathbf{x}_k$ assuming a nominal head width and orientation. We form the stacked coordinate vector
\begin{equation}
\mathbf{x} =
[\mathbf{x}_1^{\mathsf{T}},\dots,\mathbf{x}_K^{\mathsf{T}}]^{\mathsf{T}}
\in \mathbb{R}^{Kd},
\label{eq:coord_vector}
\end{equation}
and use it as the conditioning input to the neural filter generators.

Woofer and tweeter filters are generated by two separate neural networks trained independently:
\begin{equation}
\mathbf{g}^{(w)} = f_{\theta_w}(\mathbf{x}), \qquad
\mathbf{g}^{(t)} = f_{\theta_t}(\mathbf{x}),
\label{eq:two_models}
\end{equation}
where $f_{\theta_w}(\cdot)$ outputs the stacked woofer filter vector and $f_{\theta_t}(\cdot)$ outputs the stacked tweeter filter vector. This separation matches practical differences between bands (e.g., filter lengths and regularization strength) and simplifies training and deployment. 

\subsection{Baseline PSZ objective}
\label{subsec:baseline_loss}

For each band $b\in\{w,t\}$, we adopt the PSZ pretraining objective used in the Binaural Spatially Adaptive Neural Network (BSANN) framework \cite{jiang2026bsann}, which serves as our baseline loss. The woofer model $f_{\theta_w}$ and tweeter model $f_{\theta_t}$ are trained independently with band-specific losses of identical form:
\begin{equation}
L^{(b)}_{\mathrm{psz}}
=
\alpha L^{(b)}_{\mathrm{BZ}}
+
(1-\alpha) L^{(b)}_{\mathrm{DZ}}
+
\beta L^{(b)}_{\mathrm{gain}}
+
\gamma L^{(b)}_{\mathrm{compact}}.
\label{eq:Lpsz}
\end{equation}

Here, $L^{(b)}_{\mathrm{BZ}}$ and $L^{(b)}_{\mathrm{DZ}}$ are mean-squared-error (MSE) terms governing acoustic reproduction. Specifically, $L^{(b)}_{\mathrm{BZ}}$ enforces magnitude matching to a target response at the bright-zone control points, while $L^{(b)}_{\mathrm{DZ}}$ minimizes the reproduced energy at the dark-zone control points. Their relative trade-off is controlled by $\alpha\in[0,1]$. 
The remaining terms, $L^{(b)}_{\mathrm{gain}}$ and $L^{(b)}_{\mathrm{compact}}$, regularize the predicted filters in the frequency and time domains, respectively. $L^{(b)}_{\mathrm{gain}}$ penalizes frequency-domain magnitudes that exceed a band-specific limit $g^{(b)}_{\max}$ to prevent excessive control effort, while $L^{(b)}_{\mathrm{compact}}$ penalizes late-time filter energy using a weighted window to encourage compact temporal responses. The corresponding penalty strengths are controlled by $\beta\ge 0$ and $\gamma\ge 0$. 

Overall, $\alpha$, $\beta$, and $\gamma$ weight the four terms in a weighted-sum surrogate for a multi-objective design problem (bright-zone reproduction, dark-zone leakage suppression, and regularization). Since the constituent terms are defined in different domains and may have different physical units, the aggregate loss $L^{(b)}_{\mathrm{psz}}$ is not intended to be interpreted as a single physically commensurate energy functional. For numerical conditioning, we apply fixed scale factors (scale balancing) so that the major terms have comparable magnitudes early in training: $L^{(b)}_{\mathrm{BZ}}$ and $L^{(b)}_{\mathrm{DZ}}$ are multiplied by $10^{3}$ and $L^{(b)}_{\mathrm{compact}}$ is multiplied by $5$. These fixed multipliers can be absorbed into the effective weights and are used only for numerical conditioning. For brevity, the explicit integral and summation expressions for these four constituent terms are omitted here, as they are formulated exactly as in our previous work \cite{jiang2026bsann}. Following this prior work, we adopt the hyperparameter assignment $\alpha=0.5$, $\beta=0.5$, and $\gamma=0.5$ for both filter bands.

\subsection{Neighbor-consistency regularization}
\label{subsec:nc_loss}

The baseline PSZ objective in \eqref{eq:Lpsz} does not explicitly constrain how the predicted filters vary with the input coordinates. As a result, small coordinate perturbations may lead to disproportionately large changes in the generated filters. To reduce this sensitivity, we regularize each band-specific mapping $f_{\theta_b}$ so that nearby input coordinates produce similar filter coefficients.

For each training sample with nominal coordinate vector $\mathbf{x}\in\mathbb{R}^{Kd}$, we draw an elementwise random perturbation
\begin{equation}
\boldsymbol{\delta} \sim \mathcal{U}\!\left([-\Delta,\,\Delta]^{Kd}\right),
\label{eq:delta_uniform}
\end{equation}
and form a perturbed coordinate
\begin{equation}
\mathbf{x}^{\prime} = \Pi_{\mathcal{B}}\!\left(\mathbf{x} + \boldsymbol{\delta}\right),
\label{eq:x_pert_clip}
\end{equation}
where $\Pi_{\mathcal{B}}(\cdot)$ clips each coordinate to the valid training bounds $\mathcal{B}$. The predicted band-$b$ filter vectors are
\begin{equation}
\mathbf{g}^{(b)} = f_{\theta_b}(\mathbf{x}), \qquad
\mathbf{g}^{(b)\prime} = f_{\theta_b}(\mathbf{x}^{\prime}).
\label{eq:g_and_gpert}
\end{equation}

In our two-zone setting, the pipeline distinguishes between non-overlapping and overlapping zone configurations. Geometrically, overlap occurs when the Euclidean distance between the listener center coordinates falls below a predefined threshold $d_{\mathrm{ov}}$. In the overlapping regime, dark-zone suppression becomes ill-defined in the shared region because the same physical space would be simultaneously required to be bright and dark for the same program. Accordingly, we disable the dark-zone term in the overlapping regime by masking the dark-zone transfer functions with an overlap indicator (i.e., the dark-zone loss is evaluated only when $\|\mathbf{x}_1-\mathbf{x}_2\|_2 > d_{\mathrm{ov}}$). Because this threshold introduces a regime discontinuity, we apply neighbor consistency only when the original and perturbed coordinates remain in the same regime.

Let
\begin{equation}
r(\mathbf{x})=\mathbb{I}
\!\left(\left\|\mathbf{x}_1-\mathbf{x}_2\right\|_2 > d_{\mathrm{ov}}\right),
\label{eq:region_indicator}
\end{equation}
where $\mathbb{I}(\cdot)$ denotes the indicator function, which equals 1 if its argument is true and 0 otherwise, $d_{\mathrm{ov}}$ is the overlap threshold, and $\mathbf{x}_1,\mathbf{x}_2 \in \mathbb{R}^{d}$ denote the respective center-of-head coordinates of the two listeners. Thus, $r(\mathbf{x})=1$ denotes the non-overlapping regime, and $r(\mathbf{x})=0$ denotes the overlapping regime. We define the same-region mask
\begin{equation}
m(\mathbf{x},\mathbf{x}^{\prime})=\mathbb{I}\!\left(r(\mathbf{x})=r(\mathbf{x}^{\prime})\right).
\label{eq:same_region_mask}
\end{equation}

Using this mask, the neighbor-consistency loss for band $b\in\{w,t\}$ is defined as the masked mean-squared difference between filter vectors:
\begin{equation}
L_{\mathrm{nc}}^{(b)}(\theta_b)
=
\frac{
\mathbb{E}\!\left[
m(\mathbf{x},\mathbf{x}^{\prime})\,
\frac{1}{D^{(b)}}\left\|
f_{\theta_b}(\mathbf{x})-f_{\theta_b}(\mathbf{x}^{\prime})
\right\|_2^2
\right]
}{
\mathbb{E}\!\left[m(\mathbf{x},\mathbf{x}^{\prime})\right]+\varepsilon
},
\label{eq:nc_loss_masked}
\end{equation}
where $D^{(b)}$ is the number of filter coefficients in $\mathbf{g}^{(b)}$ and $\varepsilon$ is a small constant. In minibatch training, \eqref{eq:nc_loss_masked} is approximated by computing the mean-squared error (MSE) of the filter coefficients per sample and averaging only over samples satisfying the same-region condition. During training, the overlap-aware masking is primarily used to keep the objective well-defined when a sampled anchor or its perturbed neighbor enters the overlap regime.
In Sections~\ref{sec:sim} and \ref{sec:meas}, we restrict evaluation anchors to non-overlapping configurations, so the overlap-aware masking is not triggered during evaluation and does not affect the reported metrics.

This regularization reduces the sensitivity of the coordinate-to-filter mapping to coordinate estimation errors. The perturbation range $\Delta$ and the loss weight $\lambda_b$ (Section~\ref{subsec:overall_objective}) control the robustness--quality trade-off. 

In training, the perturbation in \eqref{eq:delta_uniform} is applied to the stacked coordinate input $\mathbf{x}\in\mathbb{R}^{Kd}$ for symmetry across listeners. In evaluation (Sections~\ref{sec:sim} and \ref{sec:meas}), we perturb only Listener 2 to obtain a controlled and interpretable robustness analysis; extending the perturbation to both listeners is straightforward but is left for future work.

\subsection{Overall training objective}
\label{subsec:overall_objective}

Each band-specific model is trained independently by combining the baseline PSZ objective with neighbor consistency:
\begin{equation}
L^{(b)}(\theta_b) =
L^{(b)}_{\mathrm{psz}}(\theta_b) + \lambda_b\,L^{(b)}_{\mathrm{nc}}(\theta_b),
\qquad b\in\{w,t\},
\label{eq:total_loss_band}
\end{equation}
where $\lambda_b \ge 0$ controls the stability--quality trade-off in band $b$. For numerical conditioning, we apply a fixed scale factor of $10^{3}$ to $L^{(b)}_{\mathrm{nc}}$ during training; this multiplier can be absorbed into the effective weight $\lambda_b$ and is used only for numerical conditioning. Larger $\lambda_b$ typically improves robustness to coordinate uncertainty but may reduce nominal performance if over-regularized. The perturbation range $\Delta$ is chosen to match the spatial scales relevant to localization uncertainty.

%%%%%%%%%%%%%%%%%%%%%%%%%%%%%%%%%%%%%%%%%%%%%%%%%%%%%%%%%%%%%%
%%%%%%%%%%%%%%%%%%%%     Metrics     %%%%%%%%%%%%%%%%%%%%%%%
%%%%%%%%%%%%%%%%%%%%%%%%%%%%%%%%%%%%%%%%%%%%%%%%%%%%%%%%%%%%%%

\section{Evaluation Protocol and Metrics}
\label{sec:eval}

This section describes the evaluation protocol designed to isolate robustness to localization uncertainty, and the metrics used to quantify both PSZ isolation performance and spatial stability under coordinate perturbations.

\subsection{Decoupled evaluation under localization uncertainty}
\label{subsec:decoupled_eval}

Let $\mathbf{x}=[\mathbf{x}_1^{\mathsf{T}},\mathbf{x}_2^{\mathsf{T}}]^{\mathsf{T}}\in\mathbb{R}^{Kd}$ denote the physical listener coordinates (with $K=2$ and typically $d=2$). In practical systems, the filter generator receives an estimated coordinate vector
\begin{equation}
\hat{\mathbf{x}} = \mathbf{x} + \boldsymbol{\epsilon},
\label{eq:coord_noise_model}
\end{equation}
where $\boldsymbol{\epsilon}$ models localization or tracking error.

The key observation motivating this protocol is that a 
listener can be physically stationary while $\hat{\mathbf{x}}$ 
fluctuates due to tracking noise. To evaluate robustness to this uncertainty without conflating it with true listener motion, we decouple the physical acoustics used for evaluation from the noisy coordinate conditioning used for filter generation. Specifically, for each physical configuration $\mathbf{x}$, we keep the acoustic transfer functions fixed at $\mathbf{x}$ and generate filters using perturbed input coordinates $\hat{\mathbf{x}}$:
\begin{equation}
\mathbf{g}^{(w)} = f_{\theta_w}(\hat{\mathbf{x}}), \qquad
\mathbf{g}^{(t)} = f_{\theta_t}(\hat{\mathbf{x}}),
\label{eq:eval_filters_from_noisy_coords}
\end{equation}
while all performance metrics are computed using the ATFs corresponding to the same fixed physical configuration $\mathbf{x}$. For a controlled and interpretable robustness evaluation, we perturb only the coordinates of Listener~2 while keeping Listener~1 fixed; extending the perturbations to both listeners is straightforward but is not pursued here.

\subsection{Coordinate neighborhoods}
\label{subsec:neighborhoods}

For a physical anchor $\mathbf{x}$, we define a neighborhood of perturbed coordinate inputs
\begin{equation}
\mathcal{P}(\mathbf{x})=\{\hat{\mathbf{x}}^{(i)}\}_{i=1}^{N},
\label{eq:neighborhood_set}
\end{equation}
where each $\hat{\mathbf{x}}^{(i)}$ is obtained by perturbing the relevant listener coordinates and clipping to the valid bounds, consistent with Section~\ref{subsec:nc_loss}. The specific construction of $\mathcal{P}(\mathbf{x})$ depends on the experimental setting and is described in Sections~\ref{sec:sim} and~\ref{sec:meas}.

\subsection{Sound-zone separation metrics}
\label{subsec:izi_ipi}

We quantify sound-zone separation using inter-zone isolation (IZI) and inter-program isolation (IPI), defined in \cite{qiao2022metrics}, and  evaluate these metrics frequency-wise. To evaluate the robustness against localization uncertainty, we employ a decoupled evaluation protocol: the reproduced pressures are computed using the ATFs at the true physical positions $\mathbf{x}$, while the filters are generated based on the estimated (perturbed) coordinates $\hat{\mathbf{x}}$.

Let $\mathbf{P}^{(j)}_{k,c}(\omega; \mathbf{x}, \hat{\mathbf{x}}) \in \mathbb{C}^{N_e}$ denote the complex reproduced pressure vector at the $N_e$ control points of ear $c\in\{L,R\}$ of listener $k$ when only program $j$ is active. This is obtained by applying the filters generated from $\hat{\mathbf{x}}$ in \eqref{eq:eval_filters_from_noisy_coords} to the ground-truth ATFs at the fixed physical configuration $\mathbf{x}$. We then stack both ears as
\begin{equation}
\mathbf{P}^{(j)}_{k}(\omega; \mathbf{x}, \hat{\mathbf{x}})
=
\big[\mathbf{P}^{(j)\mathsf{T}}_{k,L}(\omega; \mathbf{x}, \hat{\mathbf{x}}),\ 
\mathbf{P}^{(j)\mathsf{T}}_{k,R}(\omega; \mathbf{x}, \hat{\mathbf{x}})\big]^{\mathsf{T}}
\in \mathbb{C}^{N_k},
\end{equation}
so that $N_k=2N_e$. For each frequency $\omega$, we define the target energy at listener $k$ as
\begin{equation}
E^{\mathrm{tar}}_{k}(\omega) = \|\mathbf{P}^{(k)}_{k}(\omega; \mathbf{x}, \hat{\mathbf{x}})\|_2^2,
\label{eq:etar}
\end{equation}
the interference energy at listener $k$ from other programs as
\begin{equation}
E^{\mathrm{int}}_{k}(\omega) = \sum_{j\neq k} \|\mathbf{P}^{(j)}_{k}(\omega; \mathbf{x}, \hat{\mathbf{x}})\|_2^2,
\label{eq:eint}
\end{equation}
and the leakage energy of program $k$ into other listener zones as
\begin{equation}
E^{\mathrm{leak}}_{k}(\omega) = \sum_{i\neq k} \|\mathbf{P}^{(k)}_{i}(\omega; \mathbf{x}, \hat{\mathbf{x}})\|_2^2.
\label{eq:eleak}
\end{equation}

The frequency-dependent IZI and IPI for listener $k$ are then defined as
\begin{equation}
\mathrm{IZI}_{k}(\omega) = 10\log_{10} \left( \frac{E^{\mathrm{tar}}_{k}(\omega)}{E^{\mathrm{leak}}_{k}(\omega)+\varepsilon} \right),
\label{eq:izi_def}
\end{equation}
\begin{equation}
\mathrm{IPI}_{k}(\omega) = 10\log_{10} \left( \frac{E^{\mathrm{tar}}_{k}(\omega)}{E^{\mathrm{int}}_{k}(\omega)+\varepsilon} \right).
\label{eq:ipi_def}
\end{equation}
Larger IZI indicates less leakage of program $k$ into other zones, while larger IPI indicates lower interference from other programs at listener $k$.

\subsection{Robustness and quality metrics}
\label{subsec:quality_metrics}

We report PSZ isolation quality using the IZI and IPI metrics defined in Section~\ref{subsec:izi_ipi}. To summarize performance under localization uncertainty, we aggregate these metrics over a set of perturbed input coordinates $\mathcal{P}(\mathbf{x}) = \{\hat{\mathbf{x}}^{(i)}\}_{i=1}^N$ centered around the true physical position $\mathbf{x}$.

For a given scalar metric $q(\mathbf{x}, \hat{\mathbf{x}})$ (representing either IZI or IPI) evaluated at each perturbed input, we report the neighborhood median and a lower-bound statistic to characterize both typical and worst-case robustness:
\begin{equation}
\begin{aligned}
    q_{\mathrm{med}} &= \mathrm{median}\{q(\mathbf{x}, \hat{\mathbf{x}}^{(i)})\}_{i=1}^{N}, \\
    q_{\mathrm{worst}} &= \operatorname{LB}\!\left(\{q(\mathbf{x}, \hat{\mathbf{x}}^{(i)})\}_{i=1}^{N}\right).
\end{aligned}
\label{eq:quality_summaries}
\end{equation}

For the sparse measurement grids where $N$ is small, we take $\operatorname{LB}(\cdot)=\min$, yielding
$q_{\mathrm{worst}} = q_{\min} = \min_i q(\mathbf{x}, \hat{\mathbf{x}}^{(i)})$.
However, for dense simulation neighborhoods, the minimum can be overly sensitive to single-sample outliers. In such cases, we set $\operatorname{LB}(\cdot)=\mathrm{CVaR}_{10}(\cdot)$ and adopt the conditional value-at-risk at 10\% ($\mathrm{CVaR}_{10}$) to provide a more stable estimate of worst-case performance. Let the evaluated metrics $\{q(\mathbf{x}, \hat{\mathbf{x}}^{(i)})\}_{i=1}^{N}$ be sorted in ascending order as $q_{(1)}\le \cdots \le q_{(N)}$. The empirical $\mathrm{CVaR}_{10}$ is calculated as the mean of the bottom 10\% of samples:
\begin{equation}
\mathrm{CVaR}_{10}(q)=\frac{1}{\lceil 0.1N\rceil}\sum_{i=1}^{\lceil 0.1N\rceil} q_{(i)}.
\label{eq:cvar10}
\end{equation}
Using $q_{\mathrm{med}}$ alongside $\mathrm{CVaR}_{10}$ allows for a comprehensive assessment of the system's reliability in the presence of coordinate estimation errors.

\subsection{Stability metrics}
\label{subsec:stability_metrics}

Nominal quality alone does not fully capture robustness to localization uncertainty. We therefore quantify spatial stability by measuring the sensitivity of performance metrics to small coordinate perturbations while keeping the physical acoustics fixed.

For a physical position $\mathbf{x}$ and its associated set of estimated coordinates $\mathcal{P}(\mathbf{x})=\{\hat{\mathbf{x}}^{(i)}\}_{i=1}^{N}$, let $q_i \triangleq q(\mathbf{x}, \hat{\mathbf{x}}^{(i)})$ denote the value of a scalar metric (e.g., IZI) evaluated at $\hat{\mathbf{x}}^{(i)}$. We define a neighborhood graph over the perturbed inputs with an edge set $\mathcal{E}\subset\{(i,j): i\neq j\}$ connecting nearby perturbations. For each edge $e=(i,j)\in\mathcal{E}$, we compute a normalized variation rate (in dB/m):
\begin{equation}
\sigma_e=\frac{|q_i-q_j|}{d_{ij}}, \qquad d_{ij}=\left\|\hat{\mathbf{x}}^{(i)}-\hat{\mathbf{x}}^{(j)}\right\|_2,
\label{eq:step_per_m}
\end{equation}
where $d_{ij}$ is the distance between the two estimated coordinates. From $\{\sigma_e\}_{e\in\mathcal{E}}$, we report the mean and root-mean-square (RMS) stability summaries (where lower values indicate superior stability):
\begin{equation}
\sigma_{\mathrm{mean}}=\frac{1}{|\mathcal{E}|}\sum_{e\in\mathcal{E}} \sigma_e, \qquad
\sigma_{\mathrm{rms}}=\sqrt{\frac{1}{|\mathcal{E}|}\sum_{e\in\mathcal{E}} \sigma_e^2}.
\label{eq:stability_summaries}
\end{equation}
These statistics provide a direct measure of the local gradient of the metric with respect to coordinate errors. While $\sigma_{\mathrm{mean}}$ characterizes the typical sensitivity across the neighborhood, $\sigma_{\mathrm{rms}}$ penalizes larger variations more heavily to capture occasional sharp performance drops.

\subsection{Aggregating across anchors and reporting improvements}
\label{subsec:aggregation}

For multi-anchor evaluations, we compute the neighborhood summaries in \eqref{eq:quality_summaries}--\eqref{eq:stability_summaries} for each physical anchor and report their distributions and averages across all anchors. When comparing a baseline model with a neighbor-consistent model, we report improvement percentages using a consistent sign convention where positive values always indicate superior performance for the proposed model. For a quality metric $q \in \{\text{IZI}, \text{IPI}\}$ (where higher is better), the improvement is defined as
\begin{equation}
\mathrm{Imp}_q(\%) = 100 \cdot \frac{q_{\mathrm{nc}}-q_{\mathrm{base}}}{|q_{\mathrm{base}}|+\varepsilon},
\label{eq:improve_quality}
\end{equation}
whereas for a stability metric $\sigma$ (where lower is better), it is defined as
\begin{equation}
\mathrm{Imp}_\sigma(\%) = 100 \cdot \frac{\sigma_{\mathrm{base}}-\sigma_{\mathrm{nc}}}{\sigma_{\mathrm{base}}+\varepsilon}.
\label{eq:improve_stability}
\end{equation}
In these expressions, the subscripts ``base'' and ``nc'' denote the baseline and neighbor-consistent models, respectively, and $\varepsilon$ is a small constant for numerical stability.

%%%%%%%%%%%%%%%%%%%%%%%%%%%%%%%%%%%%%%%%%%%%%%%%%%%%%%%%%%%%%%
%%%%%%%%%%%%%%%%%%%%   Simulation Results  %%%%%%%%%%%%%%%%%%%
%%%%%%%%%%%%%%%%%%%%%%%%%%%%%%%%%%%%%%%%%%%%%%%%%%%%%%%%%%%%%%

\section{Simulation Setup and Results}
\label{sec:sim}

This section reports numerical simulation experiments designed to evaluate robustness to coordinate uncertainty using the decoupled evaluation protocol and the metrics defined in Section~\ref{sec:eval}. In the simulation analysis, the frequency-dependent IZI and IPI metrics are first aggregated within each band using the log-mean across frequency, and the resulting scalar values are used for all reported results.

\subsection{Setup and data}
\label{subsec:sim_setup}

We evaluate the proposed neighbor-consistent training in simulation using a split-band loudspeaker system, with the woofer and tweeter models trained and evaluated independently. In this simulation setup, the center-of-head position of Listener~1 is fixed at $\mathbf{x}_1=[-0.40,\,1.10]^{\mathsf{T}}$~m. Listener~2 anchors are sampled on a 2-D horizontal plane. To avoid entering the overlap regime (Section~\ref{subsec:nc_loss}) for any perturbed input, we restrict Listener~2 anchor locations to satisfy
$\|\mathbf{x}_1-\mathbf{x}_2\|_2 > d_{\mathrm{ov}} + \sqrt{2}\,r_{\max}$,
where $r_{\max}$ is the maximum perturbation magnitude along each coordinate axis in \eqref{eq:sim_offset_grid}.

For each sampled physical anchor $\mathbf{x}=[\mathbf{x}_1^{\mathsf{T}},\mathbf{x}_2^{\mathsf{T}}]^{\mathsf{T}}$, we keep the acoustic transfer functions fixed at the true positions $\mathbf{x}$ and perturb only the estimated coordinate input $\hat{\mathbf{x}}$ used for filter generation. Specifically, we generate a set of perturbed inputs by applying a 2-D offset $\boldsymbol{\delta}=[\delta_x,\delta_y]^{\mathsf{T}}$ to the Listener~2 coordinates:
\begin{equation}
\hat{\mathbf{x}}(\boldsymbol{\delta})
=
\Big[\mathbf{x}_1^{\mathsf{T}},(\mathbf{x}_2+\boldsymbol{\delta})^{\mathsf{T}}\Big]^{\mathsf{T}}.
\label{eq:sim_noisy_input}
\end{equation}
We evaluate metrics on a square offset grid
\begin{equation}
\delta_x,\delta_y \in \{-r_{\max}, -r_{\max}+\Delta_s, \ldots, r_{\max}\},
\label{eq:sim_offset_grid}
\end{equation}
with $r_{\max}=0.10$~m and $\Delta_s=0.01$~m (yielding $N=441$ perturbed inputs per anchor).

\subsection{Neighborhood graph for stability metrics}
\label{subsec:sim_edges}

To compute the stability metrics defined in Section~\ref{subsec:stability_metrics}, we construct the edge set $\mathcal{E}$ using 4-neighbor adjacencies on the offset grid in \eqref{eq:sim_offset_grid}. For each edge connecting adjacent offsets, the coordinate displacement magnitude is $d_{ij}=\Delta_s$. We then compute the normalized variation rates $\sigma_{\mathrm{mean}}$ and $\sigma_{\mathrm{rms}}$ (in dB/m) across these edges to quantify spatial sensitivity.

\subsection{Neighbor-consistency hyperparameters}
\label{subsec:sim_hparams}

The proposed neighbor-consistency regularization introduces two key hyperparameters: the perturbation range $\Delta$ (Section~\ref{subsec:nc_loss}) and the loss weight $\lambda_b$ \eqref{eq:total_loss_band}. Baseline models correspond to $\lambda_b=0$. For the neighbor-consistent (NC) model, we set $\Delta=0.01$~m and $\lambda_b=0.75$ for both the woofer and tweeter. A sensitivity study over these parameters is provided in Appendix~\ref{app:hparam}.

\subsection{Multi-anchor evaluation protocol}
\label{subsec:sim_multi_anchor}

We perform a multi-anchor evaluation by randomly sampling $K_{\mathrm{anc}}=25$ physical anchors for Listener~2. For each anchor, we evaluate IZI/IPI quality summaries (median and $\mathrm{CVaR}_{10}$) and stability summaries ($\sigma_{\mathrm{mean}}$ and $\sigma_{\mathrm{rms}}$) over the perturbed input set. All results in this section are evaluated at Listener~2 (the perturbed listener). We report the distribution of per-anchor summaries via boxplots and the overall mean$\pm$standard deviation across anchors.

\subsection{Results: woofer band}
\label{subsec:sim_woofer}

\begin{figure*}[t]
  \centering
  \includegraphics[width=\textwidth]{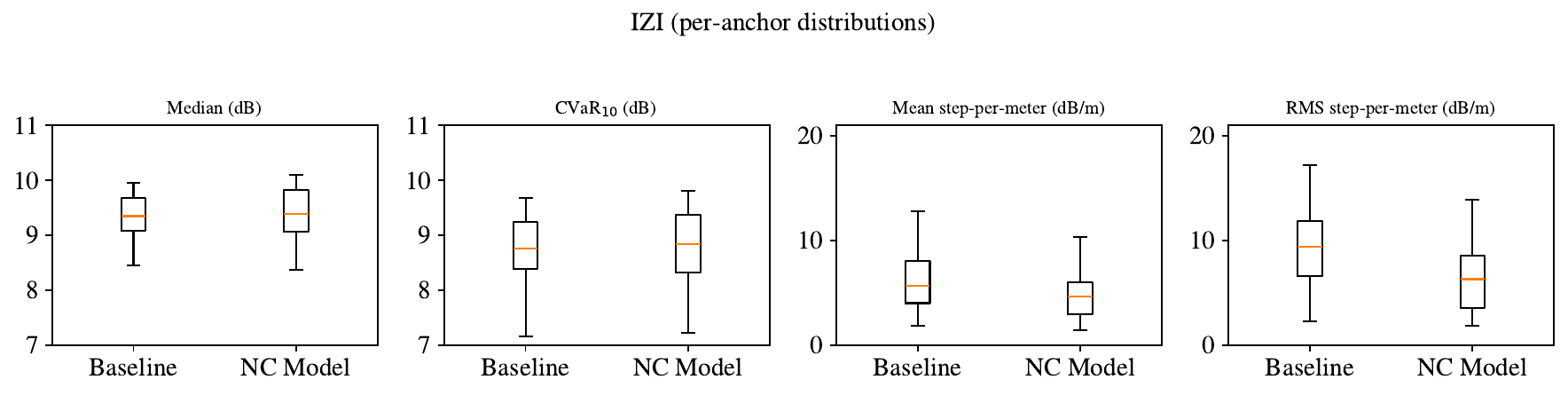}
  \includegraphics[width=\textwidth]{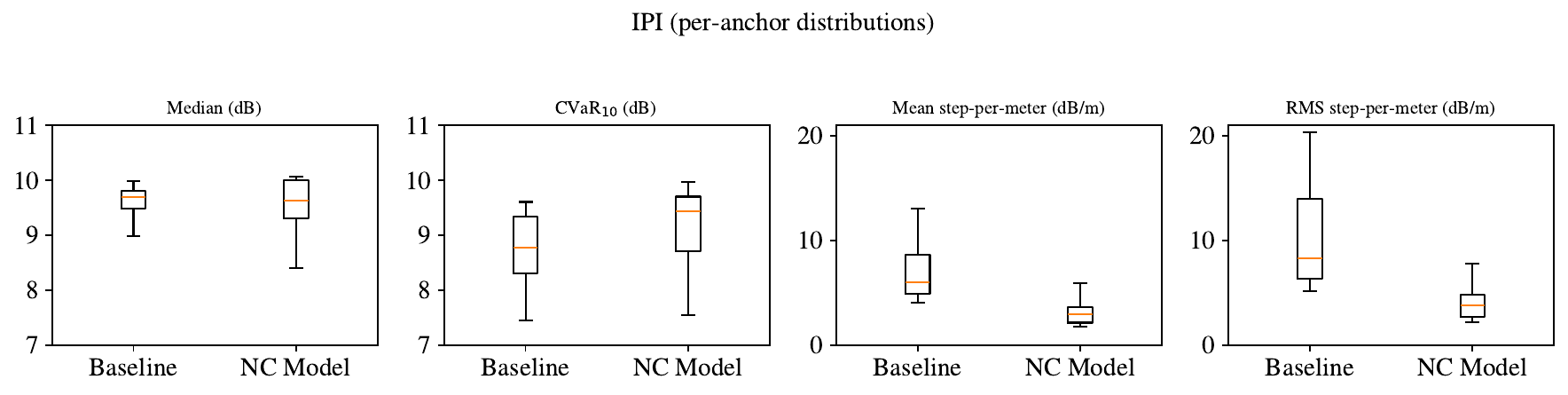}
  \caption{Simulation (woofer, Listener~2): per-anchor distributions of IZI/IPI quality summaries (median and CVaR10; higher is better) and stability summaries (mean and RMS step-per-meter; lower is better) evaluated under coordinate perturbations. Baseline corresponds to $\lambda_w=0$, while the NC model uses $\Delta=0.01$~m and $\lambda_w=0.75$.}
  \label{fig:sim_woofer_boxplots}
\end{figure*}

\begin{figure*}[t]
  \centering
  \includegraphics[width=\textwidth]{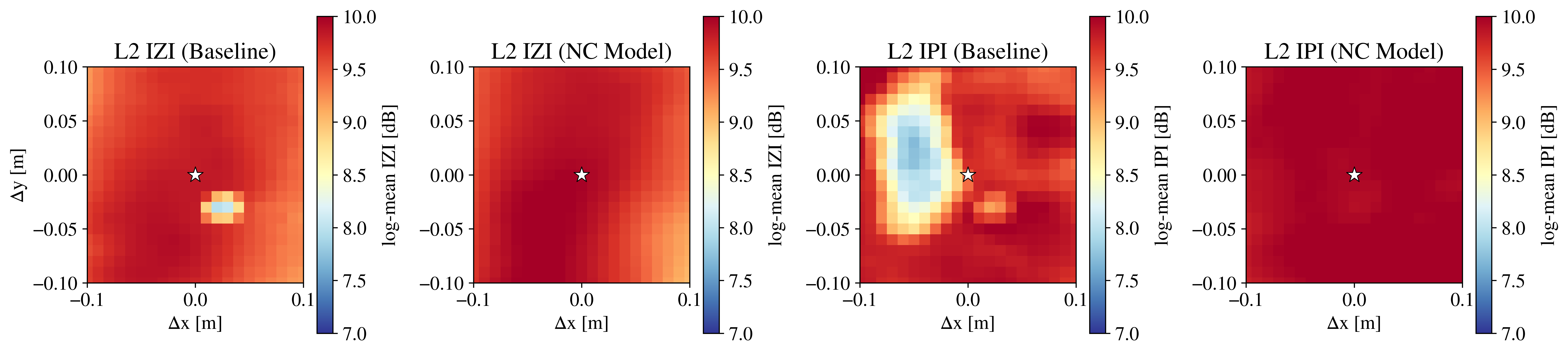}
  \caption{Simulation (woofer, Listener~2): one-anchor example of the metric landscape under coordinate perturbations. Each map plots the frequency-averaged metric value (in dB) over the offset grid; the marker indicates the unperturbed input coordinate. Baseline corresponds to $\lambda_w=0$, while the NC model uses $\Delta=0.01$~m and $\lambda_w=0.75$.}
  \label{fig:sim_woofer_one_anchor}
\end{figure*}

Fig.~\ref{fig:sim_woofer_boxplots} summarizes the per-anchor distributions of IZI/IPI neighborhood summaries for the woofer band, and Table~\ref{tab:sim_woofer_summary} reports the corresponding anchor-averaged values and relative improvements. Across anchors, the NC model largely preserves isolation quality while yielding small improvements in lower-tail performance. For IZI, the neighborhood median increases from 9.35~dB to 9.41~dB (+0.6\%) and $\mathrm{CVaR}_{10}$ increases from 8.71~dB to 8.76~dB (+0.5\%). For IPI, $\mathrm{CVaR}_{10}$ increases from 8.77~dB to 9.07~dB (+3.5\%), while the median changes only marginally (9.57~dB to 9.54~dB, $-0.3$\%).

The most consistent gains appear in the stability summaries. For IZI, the mean variation rate $\sigma_{\mathrm{mean}}$ decreases from 5.99~dB/m to 4.85~dB/m (19.1\% improvement) and the RMS variation rate $\sigma_{\mathrm{rms}}$ decreases from 9.07~dB/m to 6.63~dB/m (26.9\% improvement). For IPI, the reductions are larger: $\sigma_{\mathrm{mean}}$ decreases from 6.83~dB/m to 3.36~dB/m (50.9\% improvement) and $\sigma_{\mathrm{rms}}$ decreases from 9.92~dB/m to 4.38~dB/m (55.9\% improvement). In the woofer band, IPI exhibits larger baseline step-per-meter values than IZI, and correspondingly larger stability gains under NC regularization. The corresponding distributions in Fig.~\ref{fig:sim_woofer_boxplots} shift downward across anchors, indicating reduced local sensitivity under coordinate perturbations. This trend is also visible in the one-anchor maps in Fig.~\ref{fig:sim_woofer_one_anchor}, where the NC model exhibits a more regular metric landscape around the nominal coordinate, consistent with the reduced step-per-meter statistics.

\begin{table}[t]
\centering
\caption{Simulation (woofer, Listener~2): Anchor-averaged neighborhood summaries. Median and CVaR$_{10}$ are in dB ($\uparrow$), and stability summaries ($\sigma_{\mathrm{mean}}, \sigma_{\mathrm{rms}}$) are in dB/m ($\downarrow$). Imp.\ (\%) follows the sign convention in Section~\ref{subsec:aggregation}.}
\label{tab:sim_woofer_summary}
\small 
\begin{tabular}{llccc}
\toprule
\multicolumn{2}{c}{\textbf{Metric}} & \textbf{Baseline} & \textbf{NC Model} & \textbf{Imp. (\%)} \\ 
\midrule
\multirow{4}{*}{\textbf{IZI}} 
& Median ($\uparrow$) & 9.35 & 9.41 & +0.6\% \\
& CVaR$_{10}$ ($\uparrow$) & 8.71 & 8.76 & +0.5\% \\
& $\sigma_{\mathrm{mean}}$ ($\downarrow$) & 5.99 & 4.85 & +19.1\% \\
& $\sigma_{\mathrm{rms}}$ ($\downarrow$) & 9.07 & 6.63 & +26.9\% \\ 
\midrule
\multirow{4}{*}{\textbf{IPI}} 
& Median ($\uparrow$) & 9.57 & 9.54 & $-0.3$\% \\
& CVaR$_{10}$ ($\uparrow$) & 8.77 & 9.07 & +3.5\% \\
& $\sigma_{\mathrm{mean}}$ ($\downarrow$) & 6.83 & 3.36 & +50.9\% \\
& $\sigma_{\mathrm{rms}}$ ($\downarrow$) & 9.92 & 4.38 & +55.9\% \\ 
\bottomrule
\end{tabular}
\end{table}

\subsection{Results: tweeter band}
\label{subsec:sim_tweeter}
\begin{figure*}[t]
  \centering
  \includegraphics[width=\textwidth]{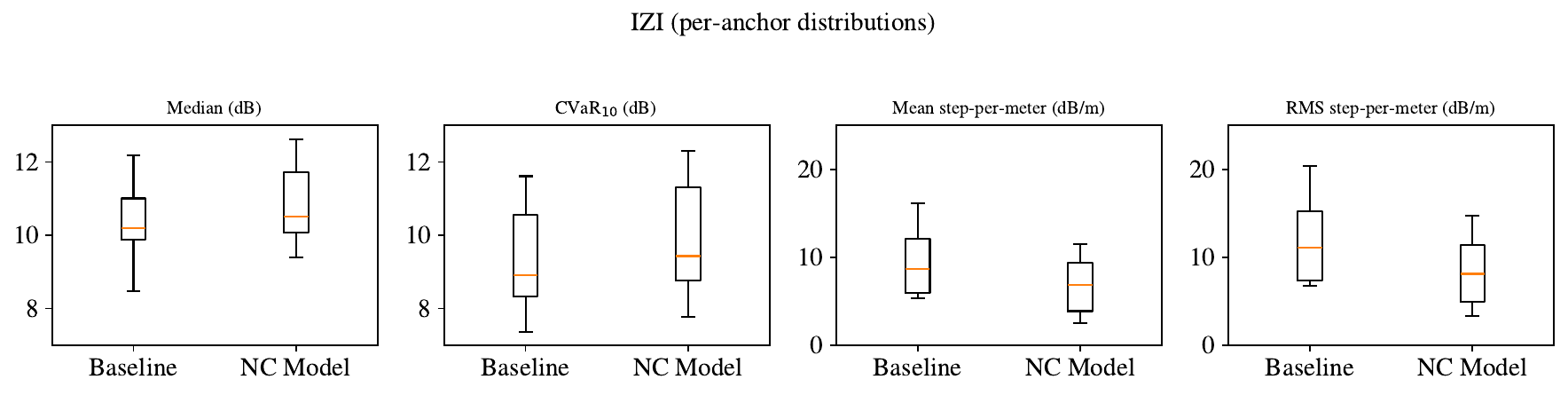}
  \includegraphics[width=\textwidth]{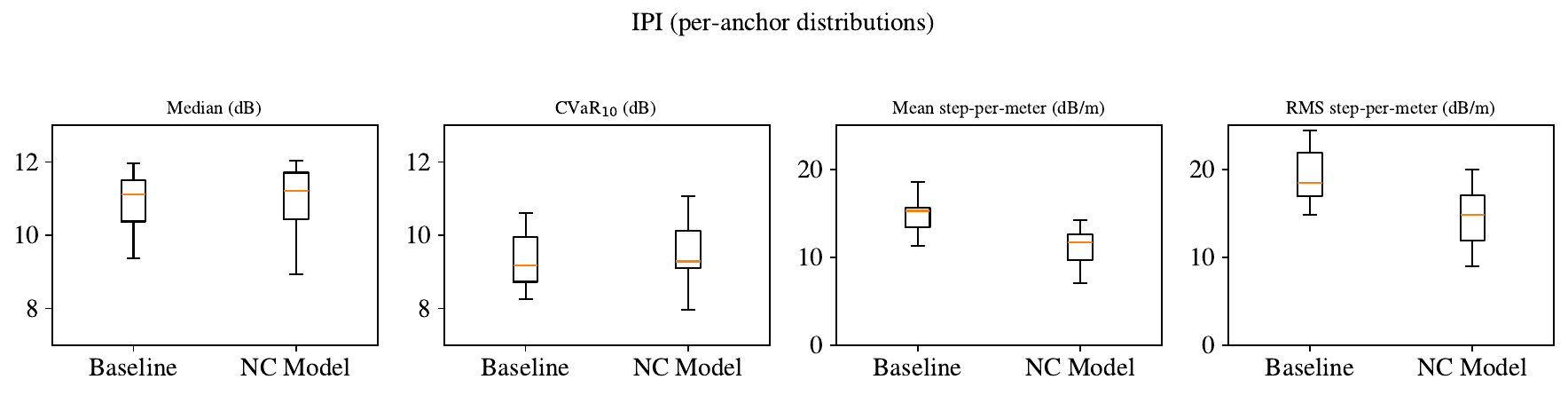}
  \caption{Simulation (tweeter, Listener~2): per-anchor distributions of IZI/IPI quality summaries (median and CVaR10; higher is better) and stability summaries (mean and RMS step-per-meter; lower is better) evaluated under coordinate perturbations. Baseline corresponds to $\lambda_t=0$, while the NC model uses $\Delta=0.01$~m and $\lambda_t=0.75$.}
  \label{fig:sim_tweeter_boxplots}
\end{figure*}

\begin{figure*}[t]
  \centering
  \includegraphics[width=\textwidth]{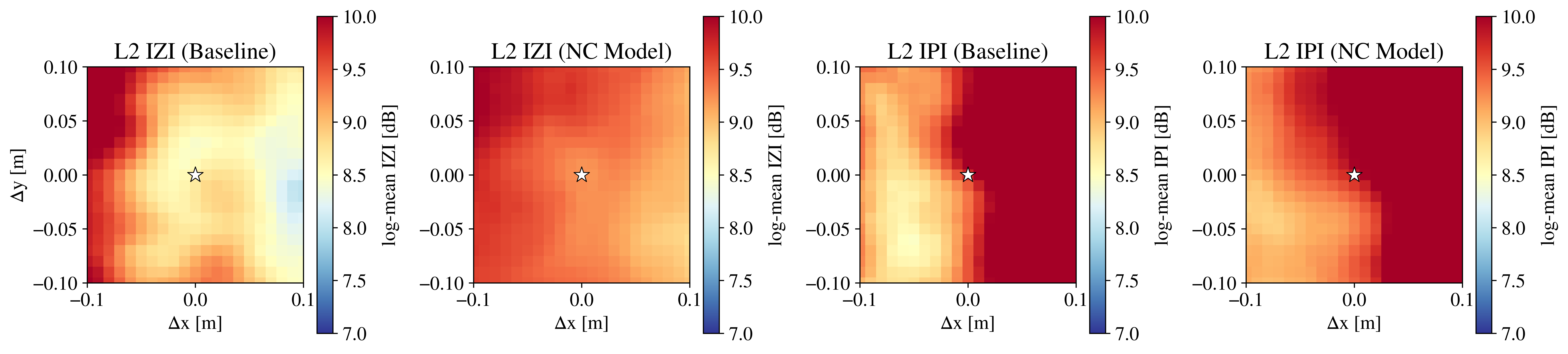}
  \caption{Simulation (tweeter, Listener~2): one-anchor example of the metric landscape under coordinate perturbations. Each map plots the frequency-averaged metric value (in dB) over the offset grid; the marker indicates the unperturbed input coordinate. Baseline corresponds to $\lambda_t=0$, while the NC model uses $\Delta=0.01$~m and $\lambda_t=0.75$.}
  \label{fig:sim_tweeter_one_anchor}
\end{figure*}

In the tweeter band (Table~\ref{tab:sim_tweeter_summary} and Fig.~\ref{fig:sim_tweeter_boxplots}), the NC model also reduces sensitivity to coordinate perturbations, as reflected by the consistently lower variation rates. For IZI, the mean variation rate $\sigma_{\mathrm{mean}}$ decreases from 9.35~dB/m to 6.65~dB/m (28.9\% improvement) and the RMS variation rate $\sigma_{\mathrm{rms}}$ decreases from 11.89~dB/m to 8.28~dB/m (30.3\% improvement). For IPI, stability improvements are similarly substantial, with $\sigma_{\mathrm{mean}}$ decreasing from 14.73~dB/m to 11.07~dB/m (24.8\% improvement) and $\sigma_{\mathrm{rms}}$ decreasing from 19.32~dB/m to 14.68~dB/m (24.0\% improvement). The visualization in Fig.~\ref{fig:sim_tweeter_one_anchor} confirms that the NC model yields a more spatially regular landscape around the unperturbed coordinate. The more noticeable quality gains in the tweeter band suggest that NC regularization helps the network learn a smoother and more robust solution in the presence of more complex high-frequency interference patterns.

\begin{table}[t]
\centering
\caption{Simulation (tweeter, Listener~2): Anchor-averaged neighborhood summaries. Median and CVaR$_{10}$ are in dB ($\uparrow$), and stability summaries ($\sigma_{\mathrm{mean}}, \sigma_{\mathrm{rms}}$) are in dB/m ($\downarrow$).}
\label{tab:sim_tweeter_summary}
\small
\begin{tabular}{llccc}
\toprule
\multicolumn{2}{c}{\textbf{Metric}} & \textbf{Baseline} & \textbf{NC Model} & \textbf{Imp. (\%)} \\
\midrule
\multirow{4}{*}{\textbf{IZI}}
& Median ($\uparrow$) & 10.42 & 10.78 & +3.5\% \\
& CVaR$_{10}$ ($\uparrow$) & 9.36 & 9.89 & +5.6\% \\
& $\sigma_{\mathrm{mean}}$ ($\downarrow$) & 9.35 & 6.65 & +28.9\% \\
& $\sigma_{\mathrm{rms}}$ ($\downarrow$) & 11.89 & 8.28 & +30.3\% \\
\midrule
\multirow{4}{*}{\textbf{IPI}}
& Median ($\uparrow$) & 10.90 & 11.02 & +1.1\% \\
& CVaR$_{10}$ ($\uparrow$) & 9.30 & 9.51 & +2.2\% \\
& $\sigma_{\mathrm{mean}}$ ($\downarrow$) & 14.73 & 11.07 & +24.8\% \\
& $\sigma_{\mathrm{rms}}$ ($\downarrow$) & 19.32 & 14.68 & +24.0\% \\
\bottomrule
\end{tabular}
\end{table}

\subsection{Discussion}
\label{subsec:sim_discussion}

The simulation results confirm that neighbor-consistency regularization improves robustness to coordinate uncertainty primarily by reducing the spatial variation rates ($\sigma_{\mathrm{mean}}$, $\sigma_{\mathrm{rms}}$) for both IZI and IPI. Across anchors, neighborhood quality summaries are preserved or improved relative to the baseline, with the most consistent gains appearing in the stability statistics. These trends align with the objective of improving spatial stability under coordinate perturbations rather than merely maximizing single-point isolation.

%%%%%%%%%%%%%%%%%%%%%%%%%%%%%%%%%%%%%%%%%%%%%%%%%%%%%%%%%%%%%%
%%%%%%%%%%%%%%%%%%%%   Measurements   %%%%%%%%%%%%%%%%%%%%%%%%
%%%%%%%%%%%%%%%%%%%%%%%%%%%%%%%%%%%%%%%%%%%%%%%%%%%%%%%%%%%%%%

\section{Experimental Validation}
\label{sec:meas}

We validate the proposed neighbor-consistency regularization using in-situ acoustic measurements. Following the decoupled evaluation principle, we test robustness by fixing physical listener locations while perturbing the estimated coordinate input to the filter generators.

\subsection{Measurement setup}
\label{subsec:meas_setup}

\begin{figure}[t]
  \centering
  \includegraphics[width=\linewidth]{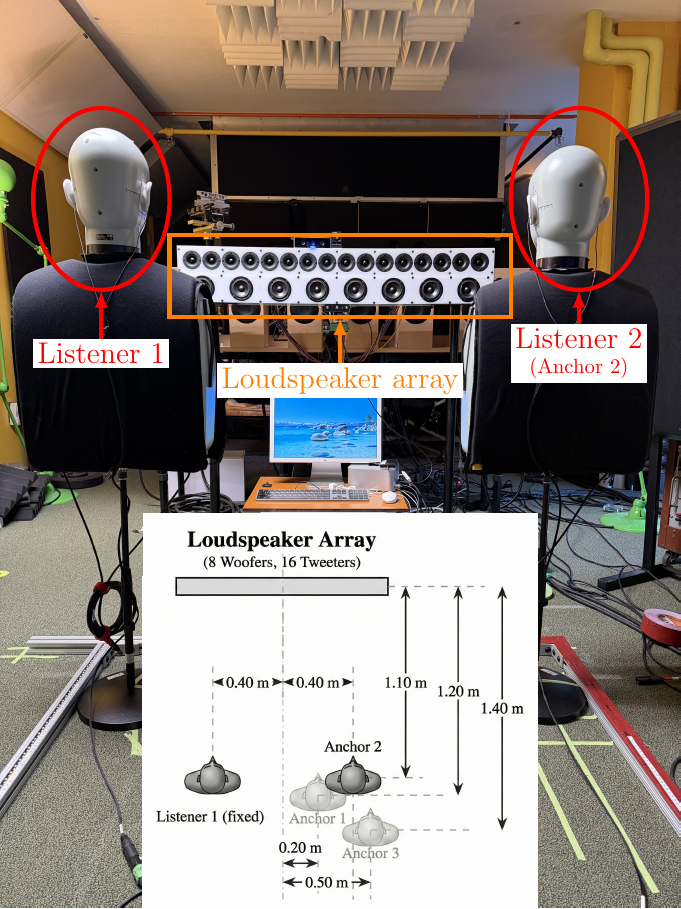}
  \caption{In-situ measurement setup. (Top) Photograph of the listening room with two static Br\"uel \& Kj\ae r head-and-torso simulators (HATS) positioned in front of a 24-driver split-band loudspeaker array. (Bottom) Top-down schematic showing the fixed position of Listener~1 and the three evaluation anchors for Listener~2.}
  \label{fig:meas_setup}
\end{figure}

The measurement setup, illustrated in Fig.~\ref{fig:meas_setup}, employs a custom-built split-band loudspeaker array facing two static Br\"uel \& Kj\ae r head-and-torso simulators (HATS). The loudspeaker system consists of 24 independently driven transducers mounted on a rigid baffle at listener height and arranged in two linear horizontal rows. The lower row contains 8 woofers (driver diameter: 0.064~m) operating over 100--2000~Hz, and the upper row contains 16 tweeters (driver diameter: 0.025~m) covering 2--20~kHz. A digital crossover at 2~kHz routes the band-specific outputs of the woofer and tweeter filter generators to the corresponding driver groups. All drivers face toward the listeners and are uniformly spaced along a horizontal aperture (overall length: 1.219~m; center-to-center spacing: 0.152~m for woofers and 0.076~m for tweeters), matching the array geometry assumed during training. Reproduced signals are captured using free-field-equalized BACCH-BM Pro in-ear microphones positioned at the ear canals of the HATS.

\subsection{Anchor configurations and coordinate perturbations}
\label{subsec:meas_protocol}

Listener~1 is fixed at $\mathbf{x}_1=[-0.40,\,1.10]^{\mathsf{T}}$~m. Listener~2 is evaluated at three anchors $\mathbf{x}_2^{(a)}$ representing different spatial configurations (Fig.~\ref{fig:meas_setup}). To emulate tracking noise, we perturb the Listener~2 input over a $3\times3$ grid $\hat{\mathbf{x}}^{(a)}(u,v;\Delta_s)$ with spacings $\Delta_s \in \{0.02,\,0.05,\,0.10\}$~m, yielding $N=9$ perturbed inputs per condition. All anchors and their perturbed inputs satisfy the non-overlapping condition $\|\mathbf{x}_1-\hat{\mathbf{x}}_2\|_2 > d_{\mathrm{ov}}$.

\subsection{Signal rendering and performance characterization}
\label{subsec:meas_processing}

For each perturbed coordinate input $\hat{\mathbf{x}}$, we query the band-specific generators to obtain the filter sets $\mathbf{g}^{(w)}$ and $\mathbf{g}^{(t)}$. Although the filters are generated in separate bands, the evaluation is conducted over the full 100~Hz to 20~kHz bandwidth simultaneously. A broadband swept-sine probe signal is filtered by both filter sets and rendered through the split-band array. The resulting pressures at each ear are then captured to compute the frequency-dependent IZI and IPI metrics. To characterize the overall system performance under realistic listening conditions, the final reported values are obtained by log-mean averaging the frequency-wise metrics across the entire evaluation band. To assess measurement repeatability, each coordinate-perturbation condition was measured three times by repeating the swept-sine playback and ear-signal recording procedure. Across repeated runs, the standard deviation of the full-band log-mean IZI/IPI metrics was below 0.20 dB for all reported conditions. These run-to-run variations were therefore small relative to the differences between the baseline and NC models and did not affect the conclusions.

\subsection{Neighborhood statistics and stability analysis}
\label{subsec:meas_summaries}

To characterize the system behavior across each $3\times3$ perturbed coordinate set ($N=9$), we evaluate neighborhood isolation quality through two key statistics: the median ($q_{\mathrm{med}}$) and the minimum ($q_{\min}$). In the context of our sparse measurement grid, $q_{\min}$ serves as a lower-bound metric to identify potential worst-case performance drops under localization uncertainty. 

Furthermore, we conduct a spatial stability analysis using the normalized variation rates $\sigma_{\mathrm{mean}}$ and $\sigma_{\mathrm{rms}}$ introduced in Section~\ref{subsec:stability_metrics}. For the measurement grid, the neighborhood graph is constructed using 4-neighbor adjacencies (horizontal and vertical connections), where the displacement magnitude for each edge is exactly $d_{ij}=\Delta_s$. These stability statistics provide a direct measure of the local sensitivity of the generated sound zones, quantifying the smoothness of the control landscape in a real-world acoustic environment.

\subsection{Results}
\label{subsec:meas_results}

We compare the baseline model ($\lambda_b=0$) against the neighbor-consistent (NC) model ($\Delta=0.01$~m, $\lambda_b=0.75$). Table~\ref{tab:meas_summary} reports the anchor-averaged absolute values and relative improvements across three perturbation spacings: $\Delta_s \in \{0.02,\,0.05,\,0.10\}$~m.

Although coordinate perturbations are applied only to Listener 2 in the conditioning input, improvements are observed at \emph{both} listeners. This behavior is expected because the neural filter generators take the joint coordinate vector as input and output the complete set of loudspeaker filters for both programs. Consequently, reducing the local sensitivity of the coordinate-to-filter mapping stabilizes the overall filter output with respect to coordinate-input perturbations, thereby improving the spatial stability of both sound zones. This behavior is visually illustrated by the one-anchor maps in Fig.~\ref{fig:meas_one_anchor_maps}.

\subsubsection{Isolation quality}
In measurements, the NC model improves absolute neighborhood quality summaries across most conditions in Table~\ref{tab:meas_summary}, including the minimum (worst-case) metric value within the $3\times3$ neighborhood. For example, at $\Delta_s=5$~cm, the minimum IZI at Listener~2 (L2) increases by 11.9\%, and the minimum IPI at Listener~1 (L1) increases by 10.0\%. Similar gains are observed across other spacings, with the largest improvement appearing in the minimum IZI at L2 (up to 16.9\% at $\Delta_s=10$~cm). Fig.~\ref{fig:meas_one_anchor_maps} provides a complementary view for Anchor~2 at $\Delta_s = 5$~cm: while the baseline exhibits pronounced spatial variation and lower metric values away from the nominal coordinate, the NC model maintains consistently higher values across the neighborhood. Compared to simulation, where NC mainly preserved neighborhood isolation, these measurement gains suggest that NC regularization may also improve generalization under real-world acoustic conditions, yielding isolation quality improvements that outperform those observed in simulation.

\subsubsection{Spatial stability}
\begin{figure*}[t]
  \centering
  \includegraphics[width=\textwidth]{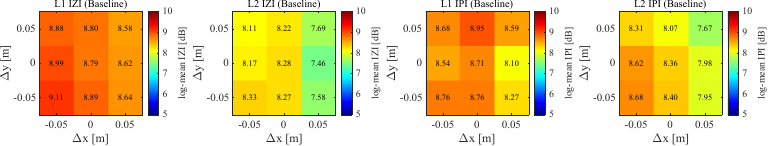}
  \vspace{0.1cm}
  
  \includegraphics[width=\textwidth]{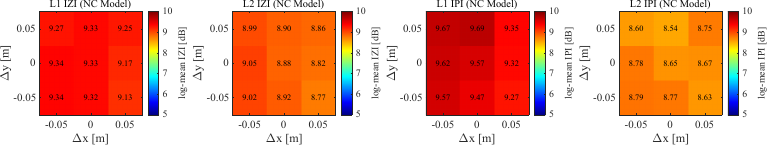}
  \caption{Measurements: One-anchor example (Anchor~2) showing $3\times3$ neighborhood maps of IZI and IPI under coordinate perturbations of $\pm5$~cm. Entries report full-band log-mean metric values (dB). Columns report L1 IZI, L2 IZI, L1 IPI, and L2 IPI; rows show Baseline (top) and NC Model (bottom).}
  \label{fig:meas_one_anchor_maps}
\end{figure*}

The stability statistics $\sigma_{\mathrm{mean}}$ and $\sigma_{\mathrm{rms}}$ show substantial reductions in local metric variability under coordinate-input perturbations. At L1, the mean variation rate $\sigma_{\mathrm{mean}}$ for IZI is reduced by up to 61.8\% at $\Delta_s=2$~cm, and the RMS variation rate $\sigma_{\mathrm{rms}}$ for IPI is reduced by up to 60.4\%. At L2, stability improvements are similarly consistent, e.g., $\sigma_{\mathrm{rms}}$ for IZI is reduced by 45.2\% at $\Delta_s=2$~cm. Across all spacings, the stability gains remain positive, with a tendency for larger relative improvements at finer perturbation spacings, consistent with the local nature of the NC regularization. The one-anchor maps in Fig.~\ref{fig:meas_one_anchor_maps} visually reflect this effect, showing a more spatially uniform metric landscape for the NC model compared to the baseline. These results confirm that NC regularization successfully smooths the control landscape, thereby reducing sharp isolation drops that could become perceptually relevant during dynamic tracking.

\begin{table*}[t]
\centering
\caption{Measurements: Anchor-averaged absolute summaries and relative improvements. Median and Min are in dB ($\uparrow$); stability summaries $\sigma_{\mathrm{mean}}$ and $\sigma_{\mathrm{rms}}$ are in dB/m ($\downarrow$). Positive percentages indicate gains.}
\label{tab:meas_summary}
\resizebox{\textwidth}{!}{%
\begin{tabular}{ll ccc ccc ccc}
\toprule
& & \multicolumn{3}{c}{\textbf{2 cm Spacing}} & \multicolumn{3}{c}{\textbf{5 cm Spacing}} & \multicolumn{3}{c}{\textbf{10 cm Spacing}} \\
\cmidrule(lr){3-5}\cmidrule(lr){6-8}\cmidrule(lr){9-11}
\textbf{Listener} & \textbf{Summary Metric} & \textbf{Baseline} & \textbf{NC Model} & \textbf{Imp. (\%)} & \textbf{Baseline} & \textbf{NC Model} & \textbf{Imp. (\%)} & \textbf{Baseline} & \textbf{NC Model} & \textbf{Imp. (\%)} \\
\midrule
\multicolumn{11}{c}{\textbf{Inter-Zone Isolation (IZI)}} \\
\midrule
\multirow{4}{*}{\textbf{L1}}
& Median ($\uparrow$) & 8.72 & 9.23 & $+5.8\%$ & 8.73 & 9.18 & $+5.1\%$ & 8.62 & 9.02 & $+4.6\%$ \\
& Min ($\uparrow$) & 8.47 & 9.13 & $+7.7\%$ & 8.24 & 8.77 & $+6.4\%$ & 8.02 & 8.49 & $+5.8\%$ \\
& $\sigma_{\mathrm{mean}}$ ($\downarrow$) & 4.61 & 1.76 & $+61.8\%$ & 5.08 & 3.14 & $+38.2\%$ & 3.94 & 3.30 & $+14.3\%$ \\
& $\sigma_{\mathrm{rms}}$ ($\downarrow$) & 6.24 & 2.39 & $+61.7\%$ & 6.85 & 4.64 & $+32.2\%$ & 5.02 & 4.34 & $+12.0\%$ \\
\cmidrule(lr){1-11}
\multirow{4}{*}{\textbf{L2}}
& Median ($\uparrow$) & 7.33 & 7.73 & $+5.5\%$ & 7.09 & 7.67 & $+8.1\%$ & 6.77 & 7.47 & $+10.2\%$ \\
& Min ($\uparrow$) & 6.75 & 7.46 & $+10.5\%$ & 6.10 & 6.83 & $+11.9\%$ & 5.36 & 6.27 & $+16.9\%$ \\
& $\sigma_{\mathrm{mean}}$ ($\downarrow$) & 8.04 & 4.68 & $+41.8\%$ & 9.76 & 6.80 & $+30.3\%$ & 8.90 & 6.21 & $+27.7\%$ \\
& $\sigma_{\mathrm{rms}}$ ($\downarrow$) & 12.08 & 6.62 & $+45.2\%$ & 13.96 & 10.49 & $+24.9\%$ & 11.31 & 8.65 & $+22.2\%$ \\
\midrule
\multicolumn{11}{c}{\textbf{Inter-Program Isolation (IPI)}} \\
\midrule
\multirow{4}{*}{\textbf{L1}}
& Median ($\uparrow$) & 8.59 & 9.32 & $+8.5\%$ & 8.64 & 9.30 & $+7.6\%$ & 8.61 & 9.27 & $+7.7\%$ \\
& Min ($\uparrow$) & 8.18 & 9.17 & $+12.1\%$ & 7.98 & 8.78 & $+10.0\%$ & 7.55 & 8.74 & $+15.8\%$ \\
& $\sigma_{\mathrm{mean}}$ ($\downarrow$) & 7.74 & 3.36 & $+56.6\%$ & 7.86 & 4.87 & $+38.0\%$ & 7.68 & 3.35 & $+57.5\%$ \\
& $\sigma_{\mathrm{rms}}$ ($\downarrow$) & 10.52 & 4.17 & $+60.4\%$ & 10.41 & 6.84 & $+34.3\%$ & 9.41 & 4.96 & $+50.7\%$ \\
\cmidrule(lr){1-11}
\multirow{4}{*}{\textbf{L2}}
& Median ($\uparrow$) & 7.40 & 7.60 & $+2.7\%$ & 7.18 & 7.48 & $+4.3\%$ & 7.11 & 7.37 & $+3.6\%$ \\
& Min ($\uparrow$) & 6.91 & 7.32 & $+5.9\%$ & 6.06 & 6.69 & $+10.3\%$ & 5.32 & 5.74 & $+8.0\%$ \\
& $\sigma_{\mathrm{mean}}$ ($\downarrow$) & 9.10 & 5.67 & $+37.7\%$ & 10.85 & 7.51 & $+30.8\%$ & 9.12 & 7.85 & $+11.6\%$ \\
& $\sigma_{\mathrm{rms}}$ ($\downarrow$) & 11.82 & 6.97 & $+41.1\%$ & 13.80 & 10.21 & $+26.0\%$ & 12.43 & 10.49 & $+13.2\%$ \\
\bottomrule
\end{tabular}
}
\end{table*}

%%%%%%%%%%%%%%%%%%%%%%%%%%%%%%%%%%%%%%%%%%%%%%%%%%%%%%%%%%%%%%
%%%%%%%%%%%%%%%%%%%%   Conclusions   %%%%%%%%%%%%%%%%%%%%%%%%%%
%%%%%%%%%%%%%%%%%%%%%%%%%%%%%%%%%%%%%%%%%%%%%%%%%%%%%%%%%%%%%%

\section{Conclusions}
\label{sec:concl}

This paper investigated the robustness of coordinate-conditioned personal sound zone (PSZ) systems under localization uncertainty by introducing neighbor-consistency (NC) regularization, which encourages neighboring input coordinates to produce similar loudspeaker filters. This approach stabilizes the reproduced sound field against fluctuations in the estimated coordinates used for filter generation.

To rigorously evaluate robustness, we introduced a decoupled evaluation protocol that isolates tracking-induced perturbations from physical listener motion. Alongside standard isolation quality metrics, we proposed spatial variation rates (reported in mean and RMS step-per-meter statistics $\sigma_{\mathrm{mean}}, \sigma_{\mathrm{rms}}$) to quantify the local sensitivity of the coordinate-to-filter mapping.

Across both simulations and in-situ measurements, NC regularization consistently and significantly reduced sensitivity to coordinate perturbations while preserving or improving isolation quality. In simulation, the proposed method yielded substantial reductions in variation rates with minimal impact on median performance. In measurements, NC regularization demonstrated even greater benefits, improving worst-case neighborhood isolation and significantly smoothing the control landscape. These results suggest that NC loss may also act as an effective regularizer that improves generalization under real-world acoustic conditions.

In sum, by smoothing the coordinate-to-filter mapping, neighbor-consistency regularization reduces abrupt degradation in local isolation and may relax the precision requirements of listener-localization systems. These properties make it a promising approach for robust, high-fidelity personal audio in practical tracking environments.

Future work will investigate the extension of this framework to dynamic tracking trajectories and conduct formal subjective listening tests to establish the perceptual correlates of the proposed stability metrics. 
%%%%%%%%%%%%%%%%%%%%%%%%%%%%%%%%%%%%%%%%%%%%%%%%%%%%%%%%%%%%%%
%%%%%%%%%%%%%%%%%%%%    Appendix%%%%%%%%%%%%%%%%%%%%%%%
%%%%%%%%%%%%%%%%%%%%%%%%%%%%%%%%%%%%%%%%%%%%%%%%%%%%%%%%%%%%%%

\appendices
\section{Sensitivity to neighbor-consistency hyperparameters}
\label{app:hparam}

This appendix reports a simulation-based sensitivity analysis of the neighbor-consistency (NC) hyperparameters: the perturbation range $\Delta$ in Section~\ref{subsec:nc_loss} and the loss weight $\lambda_b$ in \eqref{eq:total_loss_band}. Following the simulation protocol in Section~\ref{sec:sim} with $K_{\mathrm{anc}}=25$ anchors, we report anchor-averaged improvement percentages over the baseline ($\lambda_b=0$) for both isolation quality (median and CVaR$_{10}$) and spatial stability ($\sigma_{\mathrm{mean}}$ and $\sigma_{\mathrm{rms}}$ variation rates).

\begin{table*}[t]
\centering
\caption{Simulation hyperparameter sensitivity to the NC weight $\lambda_b$ with $\Delta=0.01$~m fixed. Entries report anchor-averaged improvement percentages (Imp.\ \%) over the baseline ($\lambda_b=0$) under the simulation protocol in Section~\ref{sec:sim}. Positive values indicate gains for both quality and stability, using the sign convention in Section~\ref{subsec:aggregation}.}
\label{tab:app_lambda_sweep}
\small
\setlength{\tabcolsep}{4pt}
\renewcommand{\arraystretch}{1.1}
\begin{tabular}{cc|rrrr|rrrr}
\toprule
\multirow{2}{*}{Band} & \multirow{2}{*}{$\lambda_b$}
& \multicolumn{4}{c|}{IZI Imp.\ (\%)} & \multicolumn{4}{c}{IPI Imp.\ (\%)} \\
& & Median & CVaR$_{10}$ & $\sigma_{\mathrm{mean}}$ & $\sigma_{\mathrm{rms}}$ & Median & CVaR$_{10}$ & $\sigma_{\mathrm{mean}}$ & $\sigma_{\mathrm{rms}}$ \\
\midrule
\multirow{6}{*}{Woofer}
& 0.25 & $-0.6$ & $-0.6$ & $+17.0$ & $+26.8$ & $-1.5$ & $+2.1$ & $+46.3$ & $+52.3$ \\
& 0.50 & $+0.3$ & $+0.5$ & $+21.0$ & $+29.0$ & $-2.3$ & $+1.3$ & $+49.0$ & $+54.1$ \\
& \textbf{0.75} & \textbf{+0.6} & \textbf{+0.5} & \textbf{+19.1} & \textbf{+26.9} & \textbf{\textminus 0.3} & \textbf{+3.5} & \textbf{+50.9} & \textbf{+55.9} \\
& 1.00 & $+0.2$ & $+0.9$ & $+23.4$ & $+30.5$ & $-1.0$ & $+2.3$ & $+46.9$ & $+51.4$ \\
& 1.25 & $+0.6$ & $+0.9$ & $+22.7$ & $+29.9$ & $-1.1$ & $+1.9$ & $+52.9$ & $+58.7$ \\
& 1.50 & $-0.5$ & $-0.4$ & $+28.3$ & $+35.6$ & $-1.3$ & $+1.9$ & $+50.2$ & $+54.5$ \\
\midrule
\multirow{6}{*}{Tweeter}
& 0.25 & $-0.4$ & $+0.1$ & $+10.2$ & $+11.6$ & $+1.2$ & $+1.7$ & $+20.2$ & $+18.5$ \\
& 0.50 & $+0.2$ & $-0.2$ & $+16.8$ & $+14.4$ & $+4.7$ & $+4.6$ & $+21.6$ & $+19.6$ \\
& \textbf{0.75} & \textbf{+3.5} & \textbf{+5.6} & \textbf{+28.9} & \textbf{+30.3} & \textbf{+1.1} & \textbf{+2.2} & \textbf{+24.8} & \textbf{+24.0} \\
& 1.00 & $+0.9$ & $+2.4$ & $+28.1$ & $+28.3$ & $+3.4$ & $+5.3$ & $+16.9$ & $+20.0$ \\
& 1.25 & $+1.6$ & $+2.3$ & $+28.4$ & $+30.9$ & $+2.9$ & $+3.5$ & $+18.5$ & $+18.8$ \\
& 1.50 & $-1.7$ & $+0.7$ & $+36.4$ & $+37.0$ & $+0.3$ & $+1.0$ & $+27.6$ & $+26.2$ \\
\bottomrule
\end{tabular}
\end{table*}

\begin{table*}[t]
\centering
\caption{Simulation hyperparameter sensitivity to the NC perturbation range $\Delta$ with $\lambda_b=0.75$ fixed. Entries report anchor-averaged improvement percentages (Imp.\ \%) over the baseline ($\lambda_b=0$) under the simulation protocol in Section~\ref{sec:sim}. Positive values indicate gains for both quality and stability, using the sign convention in Section~\ref{subsec:aggregation}.}
\label{tab:app_delta_sweep}
\small
\setlength{\tabcolsep}{4pt}
\renewcommand{\arraystretch}{1.1}
\begin{tabular}{cc|rrrr|rrrr}
\toprule
\multirow{2}{*}{Band} & \multirow{2}{*}{$\Delta$ (m)}
& \multicolumn{4}{c|}{IZI Imp.\ (\%)} & \multicolumn{4}{c}{IPI Imp.\ (\%)} \\
& & Median & CVaR$_{10}$ & $\sigma_{\mathrm{mean}}$ & $\sigma_{\mathrm{rms}}$ & Median & CVaR$_{10}$ & $\sigma_{\mathrm{mean}}$ & $\sigma_{\mathrm{rms}}$ \\
\midrule
\multirow{3}{*}{Woofer}
& 0.005 & $-0.9$ & $-1.3$ & $+17.4$ & $+27.7$ & $-1.5$ & $+2.6$ & $+50.4$ & $+57.3$ \\
& \textbf{0.010} & \textbf{+0.6} & \textbf{+0.5} & \textbf{+19.1} & \textbf{+26.9} & \textbf{\textminus 0.3} & \textbf{+3.5} & \textbf{+50.9} & \textbf{+55.9} \\
& 0.020 & $-2.8$ & $-1.9$ & $+34.2$ & $+38.6$ & $-4.4$ & $-1.4$ & $+53.7$ & $+58.1$ \\
\midrule
\multirow{3}{*}{Tweeter}
& 0.005 & $+0.1$ & $+0.1$ & $+0.0$ & $-2.9$ & $-2.2$ & $-0.7$ & $+9.1$ & $+12.8$ \\
& \textbf{0.010} & \textbf{+3.5} & \textbf{+5.6} & \textbf{+28.9} & \textbf{+30.3} & \textbf{+1.1} & \textbf{+2.2} & \textbf{+24.8} & \textbf{+24.0} \\
& 0.020 & $-1.5$ & $+0.2$ & $+26.0$ & $+30.2$ & $-0.5$ & $+0.2$ & $+25.7$ & $+23.6$ \\
\bottomrule
\end{tabular}
\end{table*}

Table~\ref{tab:app_lambda_sweep} sweeps $\lambda_b$ with $\Delta=0.01$~m fixed. Increasing $\lambda_b$ generally improves spatial stability (reducing $\sigma_{\mathrm{mean}}$ and $\sigma_{\mathrm{rms}}$) but eventually leads to a slight degradation in neighborhood isolation quality due to over-regularization. Across both bands, $\lambda_b=0.75$ serves as an effective knee point, yielding substantial and consistent stability gains while maintaining or even slightly improving median and CVaR$_{10}$ metrics.

Table~\ref{tab:app_delta_sweep} sweeps $\Delta$ with $\lambda_b=0.75$ fixed. A very small range ($\Delta=0.005$~m) provides limited stability benefits, particularly in the tweeter band where the control landscape is more sensitive. Conversely, a larger range ($\Delta=0.020$~m) can over-regularize the mapping, resulting in noticeable drops in neighborhood isolation. Overall, $\Delta=0.01$~m, a centimeter-scale perturbation range, provides the most balanced trade-off between local smoothness and absolute isolation performance.

Based on these results, we use $\Delta=0.01$~m and $\lambda_b=0.75$ as the default hyperparameter configuration for both bands in the main experiments.

%%%%%%%%%%%%%%%%%%%%%%%%%%%%%%%%%%%%%%%%%%%%%%%%%%%%%%%%%%%%%%
%%%%%%%%%%%%%%%%%%%%     REFERENCES     %%%%%%%%%%%%%%%%%%%%%%%
%%%%%%%%%%%%%%%%%%%%%%%%%%%%%%%%%%%%%%%%%%%%%%%%%%%%%%%%%%%%%%

\bibliographystyle{IEEEtran}
\bibliography{refs}

@article{bai2014montecarlo,
  author  = {Bai, Mingsian R. and Chen, Ching-Cheng},
  title   = {Regularization Using {Monte Carlo} Simulation to Make Optimal Beamformers Robust to System Perturbations},
  journal = {J. Acoust. Soc. Am.},
  year    = {2014},
  volume  = {135},
  number  = {5},
  pages   = {2808--2820},
  doi     = {10.1121/1.4869676}
}

@article{berkhout1993wfs,
  author  = {Berkhout, A. J. and de Vries, D. and Vogel, P.},
  title   = {Acoustic Control by Wave Field Synthesis},
  journal = {J. Acoust. Soc. Am.},
  year    = {1993},
  volume  = {93},
  number  = {5},
  pages   = {2764--2778},
doi     = {10.1121/1.405852}
}

@article{wallace2020speechprivacy,
  author  = {Wallace, Daniel and Cheer, Jordan},
  title   = {Design and evaluation of personal audio systems based on speech privacy constraints},
  journal = {J. Acoust. Soc. Am.},
  year    = {2020},
  volume  = {147},
  number  = {4},
  pages   = {2271--2282},
  doi     = {10.1121/10.0001065}
}

@article{betlehem2015psz,
  author  = {Betlehem, Terence and Zhang, Wen and Poletti, Mark A. and Abhayapala, Thushara D.},
  title   = {Personal Sound Zones: Delivering Interface-Free Audio to Multiple Listeners},
  journal = {IEEE Signal Process. Mag.},
  year    = {2015},
  volume  = {32},
  number  = {2},
  pages   = {81--91},
  doi     = {10.1109/MSP.2014.2360707}
}

@article{ahrens2010sfr,
  author  = {Ahrens, Jens and Spors, Sascha},
  title   = {Sound Field Reproduction Using Planar and Linear Arrays of Loudspeakers},
  journal = {IEEE Trans. Audio, Speech, Lang. Process.},
  year    = {2010},
  volume  = {18},
  number  = {8},
  pages   = {2038--2050},
  doi     = {10.1109/TASL.2010.2041106}
}

@article{chang2009realization,
  author  = {Chang, Ji-Ho and Lee, Chan-Hui and Park, Jin-Young and Kim, Yang-Hann},
  title   = {A Realization of Sound Focused Personal Audio System Using Acoustic Contrast Control},
  journal = {J. Acoust. Soc. Am.},
  year    = {2009},
  volume  = {125},
  number  = {4},
  pages   = {2091--2097},
  doi     = {10.1121/1.3082114}
}

@article{cheer2013carcabin,
  author  = {Cheer, Jordan and Elliott, Stephen J. and Sim{\'o}n G{\'a}lvez, Marcos Felipe},
  title   = {Design and Implementation of a Car Cabin Personal Audio System},
  journal = {J. Audio Eng. Soc.},
  year    = {2013},
  volume  = {61},
  number  = {6},
  pages   = {412--424}
}

@article{choi2002acc,
  author  = {Choi, Joung-Woo and Kim, Yang-Hann},
  title   = {Generation of an Acoustically Bright Zone with an Illuminated Region Using Multiple Sources},
  journal = {J. Acoust. Soc. Am.},
  year    = {2002},
  volume  = {111},
  number  = {4},
  pages   = {1695--1700},
  doi     = {10.1121/1.1456926}
}

@article{doclo2003robustbeamformers,
  author  = {Doclo, Simon and Moonen, Marc},
  title   = {Design of Broadband Beamformers Robust Against Gain and Phase Errors in the Microphone Array Characteristics},
  journal = {IEEE Trans. Signal Process.},
  year    = {2003},
  volume  = {51},
  number  = {10},
  pages   = {2511--2526},
  doi     = {10.1109/TSP.2003.816885}
}

@article{druyvesteyn1997personalsound,
  author  = {Druyvesteyn, W. F. and Garas, J.},
  title   = {Personal Sound},
  journal = {J. Audio Eng. Soc.},
  year    = {1997},
  volume  = {45},
  number  = {9},
  pages   = {685--701}
}

@article{elliott2012robust,
  author  = {Elliott, Stephen J. and Cheer, Jordan and Choi, Jung-Woo and Kim, Youngtae},
  title   = {Robustness and Regularization of Personal Audio Systems},
  journal = {IEEE Trans. Audio, Speech, Lang. Process.},
  year    = {2012},
  volume  = {20},
  number  = {7},
  pages   = {2123--2133},
  doi     = {10.1109/TASL.2012.2197613}
}

@inproceedings{gerzon1992metatheory,
  author    = {Gerzon, Michael A.},
  title     = {General Metatheory of Auditory Localization},
  booktitle = {Proc. Audio Eng. Soc. 92nd Conv.},
  year      = {1982},
  address   = {Vienna, Austria}
}

@misc{jiang2026bsann,
  author        = {Jiang, Hao and Choueiri, Edgar Y.},
  title         = {Stereo Audio Rendering for Personal Sound Zones Using a Binaural Spatially Adaptive Neural Network ({BSANN})},
  howpublished  = {arXiv preprint},
  year          = {2026},
  month         = jan,
  eprint        = {2601.06621},
  archivePrefix = {arXiv},
  primaryClass  = {eess.AS},
  note          = {arXiv:2601.06621},
  url           = {https://arxiv.org/abs/2601.06621}
}

@inproceedings{laine2017temporal,
  author    = {Laine, Samuli and Aila, Timo},
  title     = {Temporal Ensembling for Semi-Supervised Learning},
  booktitle = {Proc. Int. Conf. Learn. Represent. (ICLR)},
  year      = {2017}
}

@article{miyato2018vat,
  author  = {Miyato, Takeru and Maeda, Shin-ichi and Koyama, Masanori and Ishii, Shin},
  title   = {Virtual Adversarial Training: A Regularization Method for Supervised and Semi-Supervised Learning},
  journal = {IEEE Trans. Pattern Anal. Mach. Intell.},
  year    = {2019},
  volume  = {41},
  number  = {8},
  pages   = {1979--1993},
  doi     = {10.1109/TPAMI.2018.2858821}
}

@article{moles2020subband,
  author  = {Mol{\'e}s-Cases, Vicent and Pi{\~n}ero, Gema and de Diego, Mar{\'i}a and Gonzalez, Alberto},
  title   = {Personal Sound Zones by Subband Filtering and Time Domain Optimization},
  journal = {IEEE/ACM Trans. Audio, Speech, Lang. Process.},
  year    = {2020},
  volume  = {28},
  pages   = {2684--2696},
  doi     = {10.1109/TASLP.2020.3023628}
}

@article{moles2022wpmwindow,
  author  = {Mol{\'e}s-Cases, Vicent and Elliott, Stephen J. and Cheer, Jordan and Pi{\~n}ero, Gema and Gonzalez, Alberto},
  title   = {Weighted Pressure Matching with Windowed Targets for Personal Sound Zones},
  journal = {J. Acoust. Soc. Am.},
  year    = {2022},
  volume  = {151},
  number  = {1},
  pages   = {334--345},
  doi     = {10.1121/10.0009275}
}

@inproceedings{pepe2022neural,
  author    = {Pepe, Giuseppe and Gabrielli, Leonardo and Squartini, Stefano and Tripodi, Carlo and Strozzi, Nicol{\`o}},
  title     = {Digital Filters Design for Personal Sound Zones: A Neural Approach},
  booktitle = {Proc. Int. Joint Conf. Neural Netw. (IJCNN)},
  year      = {2022},
  address   = {Padua, Italy},
  doi       = {10.1109/IJCNN55064.2022.9892571}
}

@article{poletti2005spherical,
  author  = {Poletti, Mark A.},
  title   = {Three-Dimensional Surround Sound Systems Based on Spherical Harmonics},
  journal = {J. Audio Eng. Soc.},
  year    = {2005},
  volume  = {53},
  number  = {11},
  pages   = {1004--1025},
}

@article{qiao2022metrics,
  author  = {Qiao, Yue and Guadagnin, L{\'e}o and Choueiri, Edgar Y.},
  title   = {Isolation Performance Metrics for Personal Sound Zone Reproduction Systems},
  journal = {JASA Express Lett.},
  year    = {2022},
  volume  = {2},
  number  = {10},
  pages   = {104801},
  doi     = {10.1121/10.0014604}
}

@article{qiao2025sannpsz,
  author  = {Qiao, Yue and Choueiri, Edgar Y.},
  title   = {{SANN-PSZ}: Spatially Adaptive Neural Network for Head-Tracked Personal Sound Zones},
  journal = {IEEE/ACM Trans. Audio, Speech, Lang. Process.},
  year    = {2025},
  volume  = {33},
  pages   = {2735--2748},
  doi     = {10.1109/TASLPRO.2025.3581123}
}

@article{shin2014duallayer,
  author  = {Shin, Mincheol and Fazi, Filippo Maria and Nelson, Philip A. and Hirono, Fabio C.},
  title   = {Controlled Sound Field with a Dual Layer Loudspeaker Array},
  journal = {J. Sound Vib.},
  year    = {2014},
  volume  = {333},
  number = {16},
  pages   = {3794--3817},
  doi     = {10.1016/j.jsv.2014.03.025}
}

@inproceedings{sohn2020fixmatch,
  author    = {Sohn, Kihyuk and Berthelot, David and Carlini, Nicholas and Zhang, Zizhao and Zhang, Han and Raffel, Colin A. and Cubuk, Ekin Dogus and Kurakin, Alexey and Li, Chun{-}Liang},
  title     = {{FixMatch}: Simplifying Semi-Supervised Learning with Consistency and Confidence},
  booktitle = {Advances in Neural Information Processing Systems},
  volume    = {33},
  pages     = {596--608},
  year      = {2020}
}

@article{tang2025stft,
  author  = {Tang, Jun and Zhu, Wenye and Li, Xiaofei},
  title   = {Personal Sound Zones in the Short-Time {Fourier} Transform Domain with Relaxed Reverberation},
  journal = {J. Acoust. Soc. Am.},
  year    = {2025},
  volume  = {157},
  number  = {2},
  pages   = {778--796},
  doi     = {10.1121/10.0035578}
}

@inproceedings{tarvainen2017mean,
  author    = {Tarvainen, Antti and Valpola, Harri},
  title     = {Mean teachers are better role models: Weight-averaged consistency targets improve semi-supervised deep learning results},
  booktitle = {Advances in Neural Information Processing Systems},
  volume    = {30},
  year      = {2017}
}

@article{ward2001planewave,
  author  = {Ward, D. B. and Abhayapala, Thushara D.},
  title   = {Reproduction of a Plane-Wave Sound Field Using an Array of Loudspeakers},
  journal = {IEEE Trans. Speech Audio Process.},
  year    = {2001},
  volume  = {9},
  number  = {6},
  pages   = {697--707},
  doi = {10.1109/89.943347}
}

@article{wu2011multizone,
  author  = {Wu, Yan Jennifer and Abhayapala, Thushara D.},
  title   = {Spatial Multizone Soundfield Reproduction: Theory and Design},
  journal = {IEEE Trans. Audio, Speech, Lang. Process.},
  year    = {2011},
  volume  = {19},
  number  = {6},
  pages   = {1711--1720},
  doi     = {10.1109/TASL.2010.2097249}
}

@article{zhang2023cgmmrpm,
  author  = {Zhang, Junqing and Shi, Liming and Christensen, Mads Gr{\ae}sb{\o}ll and Zhang, Wen and Zhang, Lijun and Chen, Jingdong},
  title   = {{CGMM}-Based Sound Zone Generation Using Robust Pressure Matching with {ATF} Perturbation Constraints},
  journal = {IEEE/ACM Trans. Audio, Speech, Lang. Process.},
  year    = {2023},
  volume  = {31},
  pages   = {3331--3345},
  doi     = {10.1109/TASLP.2023.3306712}
}

@article{zhu2017robustaccmodel,
  author  = {Zhu, Qiaoxi and Coleman, Philip and Wu, Meng and Yang, Jun},
  title   = {Robust Acoustic Contrast Control with Reduced In-Situ Measurement by Acoustic Modelling},
  journal = {J. Audio Eng. Soc.},
  year    = {2017},
  volume  = {65},
  number  = {6},
  pages   = {460--473},
  doi     = {10.17743/jaes.2017.0016}
}

@article{zhu2017robustrepro,
  author  = {Zhu, Qiaoxi and Coleman, Philip and Wu, Meng and Yang, Jun},
  title   = {Robust Reproduction of Sound Zones with Local Sound Orientation},
  journal = {J. Acoust. Soc. Am.},
  year    = {2017},
  volume  = {142},
  number  = {1},
  pages   = {EL118--EL122},
  doi     = {10.1121/1.4994685}
}

@inproceedings{jacobsen2023living,
  author = {Jacobsen, Rune M{\o}berg and Skov, Kasper Fangel and Johansen, Stine S and Skov, Mikael B. and Kjeldskov, Jesper},
  title={Living with sound zones: A long-term field study of dynamic sound zones in a domestic context},
  booktitle = {Proc. 2023 CHI Conf. Human Factors in Computing Systems (CHI)},
  address = {New York, NY, USA},
  pages={1--14},
  year={2023}
}

@article{vindrola2021car,
  author = {Vindrola, Lucas and Melon, Manuel and Chamard, Jean-Christophe and Gazengel, Bruno},
  title   = {Use of the {Filtered-X} Least-Mean-Squares Algorithm to Adapt Personal Sound Zones in a Car Cabin},
  journal = {J. Acoust. Soc. Am.},
  volume  = {150},
  number  = {3},
  pages   = {1779--1793},
  year    = {2021},
  month   = sep,
  doi     = {10.1121/10.0005875}
}

@misc{hu2025hrtfformer,
  author        = {Hu, Xuyi and Li, Jian and Zhang, Shaojie and Goetz, Stefan and Picinali, Lorenzo and Akan, Ozgur B. and Hogg, Aidan O. T.},
  title         = {{HRTFformer}: A Spatially-Aware Transformer for Personalized {HRTF} Upsampling in Immersive Audio Rendering},
  year          = {2025},
  eprint        = {2510.01891},
  archivePrefix = {arXiv},
  primaryClass  = {cs.SD},
  doi           = {10.48550/arXiv.2510.01891},
  note          = {arXiv:2510.01891},
  url           = {https://arxiv.org/abs/2510.01891}
}

\end{document}